\documentclass[11pt,twoside]{atmp}

\usepackage{latexsym}
\usepackage{graphics}
\usepackage{amsmath}
\usepackage{amssymb}

\newcommand{\D}{\displaystyle}
\numberwithin{equation}{section}

\begin{document}
\title{A New Approach to the N-Particle Problem in QM}

\author{Joachim Schr\"oter}
\address{ Department of Physics, University of Paderborn,\\ 33098 Paderborn, Germany,\\
j@schroe.de}


 
            
\begin{abstract} In this paper the old problem of determining the
discrete spectrum of a multi-particle Hamiltonian is reconsidered. The
aim is to bring a fermionic Hamiltonian for arbitrary numbers N of
particles by analytical means into a shape such that modern numerical
methods can successfully be applied. For this purpose the
Cook-Schroeck Formalism is taken as starting point. This includes the
use of the occupation number representation. It is shown that the
N-particle Hamiltonian is determined in a canonical way by a fictional 
2-particle Hamiltonian. A special approximation of this
2-particle operator delivers an approximation of the N-particle
Hamiltonian, which is the orthogonal sum of finite dimensional
operators. A complete classification of the matrices of these
operators is given. Finally the method presented here is formulated as
a work program for practical applications. The connection with other
methods for solving the same problem is discussed.
\end{abstract}

\maketitle
\section{Introduction}
\label{intro} One of the central problems of many-particle quantum
mechanics, if not its main problem, is calculating the spectral
representation of a many-particle Hamiltonian, which typically has the
form
\begin{equation} \label{1.1} \mathbf {H}_N = \sum^{N}_{j} K_j + \frac{1}{2}
\sum^N_{j\not= k} W_{jk}\; .
\end{equation}
Here $K_j$ contains the kinetic energy of particle $j$ and the
external fields acting upon $j$, and $W_{jk}$ is the interaction of
the particles $j$ and $k$. As is well-known, this problem has a
solution if $W_{jk} = 0$. On the other hand, if $W_{jk}$ does not
vanish, the problem is ''almost'' unsolvable in a strict sense. But
the situation is not hopeless. For, what is really needed for
practical purposes, is a ''good'' approximate solution.

In this last field a tremendous work has been done, both analytically
and numerically. Its mainstreams are well-known under the labels
Thomas-Fermi method (\cite{Thom},\cite{Ferm}), Hartree-Fock method
(\cite{Hart},\cite{Fock}), density functional theory
(\cite{hoko},\cite{kosh}), configuration interaction method, Haken's
method and others. With respect to these methods and their
applications and refinements I refer e.g. to the following books
\cite{Kohan}, \cite{Ohno}, \cite{Haken}. There in addition an
abundance of papers and monographs is cited, where the methods are
also described in detail.

A common feature of these procedures is that they contain one step in
which a one-particle approximation of the N-particle problem is
carried through. With the methods of Thomas-Fermi and of Hartree-Fock
it is all, what is done. With the other methods the described first
step is followed by other ones thereby improving the accuracy of
approximation. Especially, by combining analytical and numerical
mathematics great progress is achieved. Today problems can be solved
which were regarded as unsolvable few decades ago.

Nevertheless, the question is obvious, whether there are other
approaches to a solution of the $N$-particle problem in quantum
mechanics than those mentioned above. It is the aim of this paper to
present such a new procedure. For this purpose I need some
mathematical tools which, though they are widely known, I have briefly
described in Appendix A.1. In particular the reader will find all the
notation which is used throughout the text. (More details can be found
in \cite{Cook}, \cite{Schroeck}.) The basic idea of the procedure as
well as the main results are sketched in Section 2.3.

\section{The Structure of $N$-Particle Hamiltonians}
\setcounter{equation}{0} 
{\bf 2.1:} In what follows only systems of
particles of the same kind are considered.                                           
When one starts studying a concrete sytem, its Hamiltonian is usually defined using the position-spin representation, i.e. the Hamiltonian is an operator in the Hilbert space $\bigotimes^N (L^2(\mathbb{R}^{3}) \otimes
\mathcal{S}^1)$, where $\bigotimes$ and $ \otimes$ denote tensor products, and where  $\mathcal{S}^1$ is the complex vector space of spin functions (cf. Section A.2.1). For explicit calculations this representation is very useful. But the aim of this paper is primarily a structural analysis of the Hamiltonians of a certain class of systems, and in this case a more abstract formalism is adequate. It turns out that the Cook-Schroeck formalism (cf. Appendix A.1) is very useful for this purpose.\\
Then our starting point is an arbitrary initial Hamiltonian of the shape (1.1), which is  denoted $ \bar H_N $ and defined in a Hilbert space ${\bar{ \mathcal H}}^N := \bigotimes^N {\bar {\mathcal H }^ 1}$, where $\bar{\mathcal H }^1$ is the Hilbert space of the corresponding one-particle system.\\
  Now let $K$ be the operator defined in $\bar{\mathcal
H}^1$ which contains the kinetic energy of one particle of a certain
kind together with the action of the external fields. Moreover, let
$W$ be that operator in $\bar{\mathcal H}^2$ which represents the
interaction of two particles of the kind considered. Then, using
Formula (\ref{A1.25}), $ \bar {H}_N$ defined in $ \bar{\mathcal H}^N$ is given by
\begin{equation} \label{2.1} 
     \bar H_N = \Omega_N (K) + \Omega_N (W), 
\end{equation}
where
\begin{equation}
\begin{array}{c}
\Omega_N (K) := ((N-1)!)^{-1}
\sum_{P \in {\mathcal{S}}_N} U(P) (K \otimes 1 \otimes \ldots
\otimes 1) U^\star(P),\\ [2ex]
\Omega_N (W) := (2 (N-2)!)^{-1}
\sum_{P \in {\mathcal{S}}_N} U(P) (W \otimes 1 \otimes \ldots
\otimes 1) U^\star(P) ,
\end{array}
\end{equation}
and $U(P)$ is the unitary permutation operator defined by the particle permutation P. Thus, using Formula (\ref{A1.27}), the operator $\bar H_N$
specified for Bosons or Fermions reads
\begin{equation} \label{2.2} \bar{H}^\pm_N = \Omega^\pm_N (K) +
\Omega^\pm_N(W).
\end{equation}
Here the definition $ A^\pm := S^\pm_N AS^\pm_N$ for an arbitrary operator $A$ in $\bar{\mathcal H}^N$ is applied, where $S^\pm_N $ is the symmetrizer (+) resp. the antisymmetrizer (-). Then $A^\pm$ is defined in the Hilbert space $\bar{\mathcal H}^N_\pm =  S^\pm_N [\bar
{\mathcal{H}}^N]$.\\                                    
It is well-known that the structure of $\bar{H}^\pm_N$ given
by (\ref{2.2}) is not helpful for studying its spectral problem, because the operators $\Omega^\pm_N (K)$ and
$\Omega^\pm_N(W)$ do not commute. This suggests the question if it is
possible to find an operator $T$ acting in $\bar{\mathcal H}^M,\; 1
\leq M < N$ such that
\begin{equation} \label{2.3}  \bar{H}^\pm_N = \Omega^\pm_N (T).
\end{equation}
Because the two-particle operator $ W $ cannot be
represented by a one-particle operator it holds that $M\geq 2$. If the
Hamiltonians as well as the operators $ K$ and $W$ are selfadjoint, it
turns out that $M=2$ is possible as shown by the following \\
{\bf Proposition 2.1:} Let
\begin{equation} \label{2.4} \tilde{H}_2 (\gamma) = \gamma (K \otimes
1) + \gamma (1 \otimes K) + W \;
\end{equation}
so that ${\tilde H}_2 (\gamma)$ is defined in
$ \bar{\mathcal H}^2$, and let $\gamma_0:=(N-1)^{-1}$. Then
\begin{equation} \label{2.5} \bar{H}^\pm_N = \Omega^\pm_N ({\tilde H}_2
(\gamma_0)) \; \text {and}\;\; \bar{H}^\pm_N \not= \Omega^\pm_N ({\tilde
H}_2 (\gamma)), \; \gamma \not= \gamma_0.
\end{equation}
{\bf Proof:} Using (\ref{A1.28}) yields
\begin{equation} \label{2.6}
\begin{array}{ll} \Omega^\pm_N (K) &= N S^\pm_N (K \otimes 1 \otimes
\ldots \otimes 1) S^\pm_N \\[3ex] &={\D \frac{1}{N-1} \binom N2
S^\pm_N ((K \otimes 1) \otimes \ldots \otimes 1) S^\pm_N} \\ [3ex] & {
+ \D\frac{1}{N-1} \binom N 2 S^\pm_N ((1 \otimes K) \otimes \ldots
\otimes 1) S^\pm_N} \\ [3ex] &= { \D\frac{1}{N-1} (\Omega^\pm_N (K
\otimes 1) + \Omega^\pm_N (1 \otimes K))}.
\end{array}
\end{equation}
 Since by supposition ${\tilde H}_2(\gamma), K $ and $W$
 are selfadjoint, (\ref {A1.30}) can be applied so that with the help of
(\ref{2.6}) the following relation holds:
\begin{equation}
\begin{array} {ll} \label{2.7} \Omega^\pm_N ({\tilde H}_2 (\gamma))
&\supset \gamma \Omega^\pm_N (K \otimes 1) + \gamma \Omega^\pm_N (1
\otimes K) + \Omega^\pm_N (W) \\ [2ex] &= \gamma (N-1) \Omega^\pm_N
(K) + \Omega^\pm_N (W).
\end{array}
\end{equation}
The term in the last line of (\ref{2.7}) is selfadjoint
because it is the Hamiltonian of a (possibly fictional) $N$-particle
system. Since also $\Omega^\pm_N ({\tilde H_2}(\gamma))$ is
selfadjoint (cf. Proposition A.1.11), relation (\ref{2.7}) is an equation, from which the
proposition follows immediately.\\  
{\bf 2.2:} This result is somewhat surprising. The initial Hamiltonian $\bar{H}^\pm_N$ is not determined by
${\tilde H_2} (1)$, i.e. by a Hamiltonian of a system of two particles
of the same kind, which is described by $\bar{H}^\pm_N$. Rather $\bar{H}^\pm_N$ is
determined by ${\tilde H}_2 (\gamma_0), \gamma^{-1}_0 = N-1$, which is
a two-particle Hamiltonian for particles of mass $(N-1)m_0$ and
external fields weakened by a factor $(N-1)^{-1}$, but with the same
interaction $W$ as the particles described by $\bar{H}^\pm_N$, which are
supposed to have mass $m_0$.\\
The system described by ${\tilde H}_2 (\gamma_0)$ is 
fictional.  I call it \textit {dummy system} and the operator ${\tilde
H}_2 (\gamma_0)$ \textit {dummy Hamiltonian}.\\
In Appendix A2 two simple examples are given describing dummy helium
and a solid with two dummy electrons.\\ In what follows, the operator
${\tilde H}_2 (\gamma), \gamma \not= \gamma_0$ is not needed
anymore. Therefore it is convenient to use the notation ${\tilde
H}_2 (\gamma_0)= \bar {H}_{20}$.\\ 
{\bf Corollary 2.2:} Because of (\ref{A1.31}) it
follows that
\begin{equation} \label{2.8} \bar{H}^\pm_N = \Omega^\pm_N (\bar{H}_{20}) =
\Omega^\pm_N (\bar{H}^\pm_{20})\;.
\end{equation}
{\bf 2.3:} \textit{Formula (\ref{2.8}) suggests the
basic idea of this paper: find an approximation of $\bar{H}^\pm_N$ via an
approximation of $\bar{H}^\pm_{20}$ such that the spectral problem of
$\bar{H}^\pm_N$ can be solved approximately.} \\
The details of this program are carried out for fermions in four steps, which correspond to the Chapters 3 to 6.\\ 
Chapter 3 contains a formal analysis of a restriction $H^-_N$ of the initial 
Hamiltonian $\bar{H}^-_N$. The aim is expressing the matrix elements of
$H^-_N$ in terms of the matrix elements of the restricted dummy operator
$H^-_{20}$, which is bounded. The results are summarized in Proposition 3.10.\\ 
In order
to use them for the present purposes, a properly chosen orthonormal system  $\mathcal O_1$ in the one-particle Hilbert space $\bar{\mathcal H}^1 $ is
needed. In Chapter 4 arguments are given that such a system  is obtained
via the Hartree-Fock procedure applied to $\bar{H}^-_{20}$. Thus the restrictions used in Chapter 3 can be justified. Moreover, a
heuristic argumentation suggests that $H^-_{20} $ can be "truncated"
such that an operator $ {\hat H}^-_{20} $ results, which is, depending
on a parameter $\alpha \in \mathbb{N}$, an approximation of
$H^-_{20}$.\\ 
In Chapter 5 it is shown that the operators $ {\hat
H}^-_{20} $ converge strongly to $H^-_{20}$, if $\alpha \rightarrow
\infty$. This has the consequence that the operators ${\hat H}^-_N =
{\Omega}^-_N({\hat H}^-_{20} )$ also converge strongly to
$H^-_N$. Therefore it is possible to apply the results of the theory
of spectral approximation (cf. e.g. \cite{Chat},\cite{Kato}).\\
Finally, in Chapter 6 an analysis of the operators $ {\hat H}^-_N $ is
given. It is shown that they are block-diagonal, i.e. their matrices
with respect to the chosen orthogonal basis are orthogonal sums of finite
dimensional matrices, the structure of which are analyzed in
detail. At this point numerical methods can come into play.\\ In
Chapter 7 the results of the previous chapters are summarized in the
form of a work program, which can be regarded as the main result of
this paper.
\section{The Hamiltonian $H^-_N$ and its Matrix}
\setcounter{equation}{0}
\subsection{Preliminary remarks} 
{\bf 3.1.1:} Since in what follows
only such systems are considered which consist of fermions of the same
kind, the notation introduced in Appendix A.1 can be used
throughout. The Hamiltonians of these systems usually have the
following \textit{ property. }\\ 
1.) They are unbounded, but bounded
from below.\\ 
2.) Their spectrum below a certain value $\epsilon_0$ is
discrete, otherwise continuous with possibly inserted discrete values.

Since in this paper we are only interested in the discrete spectrum
outside the continuum, i.e. in the bound states of the system, the
question is obvious, whether it is possible to restrict the spectral
problem to the discrete eigenvalues. 
In other words: is it possible to realize the following\\ 
{\bf Assumption 3.1:} There is a subspace ${\mathcal H}^1 \subset
{\bar{\mathcal H}^1} $ such that the restriction ${{H}^-_{20}}$ of
${\bar{H}^-_{20}}$ to the subspace ${\mathcal H}^2_- \subset
{\bar{\mathcal H}^2_-} $ is bounded (so that it can be defined on
${\mathcal H}^2_-$), and has the same discrete eigenvalues as
${\bar{H}^-_{20}}$ outside its continuous spectrum.\\
In this chapter it is assumed that such a subspace ${\mathcal H}^1
\subset {\bar{\mathcal H}^1} $ exists. Then $H^-_N = \Omega^-_N
(H^-_{20})$ is bounded and defined on ${\mathcal H}^N_-$. This
operator is the subject studied in the following sections. In chapter
4 arguments are given that the assumption can be realized.  \\ 
{\bf 3.1.2:} The starting point for the further considerations is the following\\
{\bf Notation 3.2:} 1.) Let ${\mathcal B}_1 = \{\phi_\kappa : \kappa \in
\mathbb{N} \}$ be an arbitrary ONB in ${\mathcal{H}}^1$, and let 
\begin{equation}  \phi_{\kappa_1 \cdots \kappa_M} :=
\phi_{\kappa_1} \otimes \cdots \otimes \phi_{\kappa_M} \in \mathcal{H}^M     
\end {equation}
with $2 \leq M \leq N$ and $\kappa_j \in \mathbb{N}, j = 1,....,M.$
Then an ONB ${\mathcal
B}^-_M \subset {\mathcal H}^M_-$ is defined by the vectors 
\begin{equation} \label{B1.10} \Psi^-_{\kappa_1 \ldots \kappa_M} :=
\sqrt{M!} S^-_M \phi_{\kappa_1 \ldots \kappa_M}
\end{equation}
with $ S^-_M$ being the antisymmetrizer (cf. (\ref{A1.6})).\\
2.)For each sequence ${\kappa_1 \cdots \kappa_M}$ of indices there is an infinite sequence $\hat{k} := (k_1, k_2, k_3,\cdots,  )$ of so called ocupation numbers $k_\kappa$ defined by
\begin{equation} \label{B1.11} k_\kappa = \sum^M_{j=1} \delta_{\kappa
\kappa_j}.
\end{equation}
Hence $k_\kappa = 1$ or $0$. Moreover there is a one-to-one correspondence
\begin{equation}
\hat{k} \longleftrightarrow {\kappa_1 \cdots \kappa_M}.
\end{equation}
Thus we can write
\begin{equation} \label{B1.13} \Psi^\pm_{\kappa_1 \ldots \kappa_{M}}
=: \Psi^\pm_M ({\hat k}).
\end{equation}
3.)The term "sequence of occupation numbers" is abbreviated by $bzf$ and the set of all $bzf$, which have exactly $M$ numbers $1$ is denoted $BZF_M$. The set $BZF$ comprises all $bzf$.
 
Now let $\langle \; \cdot,\cdot \;\rangle_2$ be the inner product in
${\mathcal H}^2_-$ and let
\begin{equation} \label{3.1} E (\hat k, \hat m): = \langle \Psi^-_2
(\hat k), H^-_{20}\Psi^-_2 (\hat m) \rangle_2
\end{equation}
be the matrix elements of the dummy Hamiltonian. Then
the matrix representation of $H^-_{20}$ reads:
\begin{equation} \label{3.2} H^-_{20} = \sum_{\hat k} \sum_{\hat m} E
(\hat k, \hat m) \Psi^-_2 (\hat k) \langle \Psi^-_2 (\hat m),\; \cdot
\;\rangle_2 \;.
\end{equation}
Using the abbreviation
\begin{equation} \label{3.3} \D{\Psi^-_2 (\hat k) \langle \Psi^-_2
(\hat m), \; \cdot\; \rangle_2 =: T (\hat k, \hat m)}
\end{equation}
together with Formula (\ref{A1.28}) yields:
\begin{equation} \label{3.4}
\begin{array} {ll} H^-_N &= \binom N2 S^-_N (H^-_{20} \otimes 1
\otimes \cdots \otimes 1) S^-_N \\ [3ex] &= \binom N2 \sum_{\hat
k} \sum_{\hat m} E ({\hat k}, {\hat m}) S^-_N (T (\hat k, \hat
m)\otimes 1 \otimes \cdots \otimes 1) S^-_N \; .
\end{array}
\end{equation} 
{\bf Remark 3.3:} Here and in what follows the sums $\sum_{\hat k}$
and $\sum_{\hat m}$ are understood to run over all $bzf$ which occur in the elements of ${\mathcal B}^-_2$. Each of these sums can be arbitrarily ordered because each
ordering of the ${\hat k}$ or the ${\hat m}$ yields an ONB. Since
$H^-_{20}$ is assumed to be bounded the sums $\sum_{\hat k}$ and
$\sum_{\hat m}$ can be interchanged.\\ {\bf Notation 3.4:} 1.)  As
usual the abbreviation
\begin{equation} \label{3.5} \langle {\hat n}' | H^-_N | {\hat n}
\rangle:= \langle \Psi^-_N ({\hat n'}), H^-_N \Psi^-_N (\hat n)
\rangle
\end{equation}
is used.\\ 2.) Let $\hat n \in {B Z F}_j$. In the
present case ${ j} = 2$ or ${ j} = N$, and in the next section also ${
j} = M$ is used with $2 \le M < N$. But irrespective of these special
choices, for each two $bzf$ an addition and a subtraction can be
defined by adding, respectively by subtracting their components. Since
these operations on two $bzf$ not necessarily result in a $bzf$ the
following notation is used (cf.\ \ref{A1.3}): "$\hat n \pm \hat m$ is
a $bzf$" or "$\hat n \pm \hat m \in {BZF}$". These expressions indicate that 
the sequence $\hat{n}\pm \hat{m}$ does not contain the numbers 2 or -1.

\subsection{The basic lemma} 
{\bf 3.2.1:} It will be shown that the
following proposition holds.\\ {\bf Lemma 3.5:} There is a function
$C$ such that for each triple $(\hat n, \hat k, \hat m)$ of $bzf$ with
$\sum_\alpha n_\alpha = N, \sum_\beta k_\beta = \sum_\beta m_\beta =
2$ the relations
\begin{equation}\label{3.6}
\begin{array}{llll} C (\hat{n}, \hat k, \hat m) &= \pm 1, &\mathit{if}
\; \hat n - \hat m \in {BZF} \; \mathit{and} \; &\hat n + \hat k -
\hat m \in {BZF}_N \\ &= 0, &\mathit{if} \; \hat n - \hat m \notin
{BZF}\; \mathit{or} \; &\hat n + \hat k - \hat m \not\in {BZF}_N \\
\end{array}
\end{equation}
hold, and that moreover
\begin{equation} \label{3.7} \D{\langle \hat n' | H^-_N | \hat n
\rangle = \sum_{\hat k} \sum_{\hat m} C ({\hat n}, {\hat k}, {\hat m})
E ({\hat k}, {\hat m}) \delta ({\hat n}', {\hat n} + {\hat k} - {\hat
m})},
\end{equation}
where $\delta$ is the Kronecker symbol and where $E
(\hat k, \hat m)$ is defined by (\ref{3.1}) (cf. also Remark 3.3).\\
From Lemma 3.5 one can draw the following\\ 
{\bf Conclusion 3.6:} If
$\hat n$ and $\hat n'$ are given, $\langle \hat n' |H^-_N| \hat n
\rangle$ can be unequal zero only if $\hat n - \hat m \in {{BZF}_
{N-2}}$ and $\hat n' - \hat k \in {{BZF}_{N-2}}$. These relations can
be satisfied only for $\binom N2$ $bzf$ $\hat m$ and $\hat k$. Hence
the sums in (\ref{3.7}) have finitely many summands.\\ 
{\bf 3.2.2:} Though the lemma is
used in this paper solely in the above version, for later purposes a
generalization of it will be proved in the next sections. (The
expenditure is the same in both cases.) In order to do so, some {\it
notation} is introduced.

Let $2 \leq M < N$ and let $A_M$ be a bounded operator defined on
${\mathcal{H}}^-_M$ such that
\begin{equation} \label{3.8} A_N = \Omega^-_N (A_M)
\end{equation}
is defined on ${\mathcal{H}}^N_-$. The matrix elements
of $A_M$ with respect of ${\mathcal B}^-_M$ are again denoted $E(\hat
k, \hat m)$ so that here
\begin{equation} \label{3.9} \sum_\alpha k_\alpha = \sum_\alpha
m_\alpha =M \; .
\end{equation}
 Moreover let
\begin{equation} \label{3.10} \langle \hat n' |A_N| \hat n \rangle: =
\langle \Psi^-_N (\hat n'), A_N \Psi^-_N (\hat n) \rangle.
\end{equation}
Finally, the function $C$ is defined as in Lemma 3.2
but with condition (\ref{3.9}).\\ {\bf Proposition 3.7:} The relation
\begin{equation} \label{3.11} \langle \hat n' | A_N | \hat n \rangle =
\sum_{\hat k} \sum_{\hat m} C(\hat n, \hat k, \hat m) E ( \hat k, \hat
m) \delta (\hat n', \hat n + \hat k - \hat m)
\end{equation}
holds. (Cf. also Remark 3.3 and Conclusion 3.6.)

\subsection{Proof of Formula (3.11)}

{\bf 3.3.1:} The starting point is Formula (\ref{A1.28}) and the
analogue to Formula (\ref{3.4}). Thus
\begin{equation} \label{3.12} \langle \hat n' | A_N| \hat n \rangle
= \binom NM \sum_{\hat k} \sum_{\hat m} E(\hat k, \hat m) Z (\hat
n', \hat n + \hat k - \hat m) ,
\end{equation}
 where
\begin{equation} \label{3.13} Z (\hat n', \hat n, \hat k, \hat m) =
\langle \Psi^-_N (\hat n'), (T (\hat k, \hat m) \otimes 1 \otimes
\cdots \otimes 1 ) \Psi^-_{N} ({\hat n}) \rangle.
\end{equation}
Here the operator $T (\hat k, \hat m)$ is defined by
strict analogy with (\ref{3.3}). Thus
\begin{equation} \label{3.14}
\begin{array}{lll} T (\hat k, \hat m) &= \Psi^-_M (\hat k) \langle
\Psi^-_M (\hat m), \;\cdot \;\rangle_M \\[3ex] &=
(M!)^{-{\frac{1}{2}}} \sum_{Q \in {\mathcal S}_M} \sigma^- (Q)
\Psi^-_M ({\hat k}) \langle \phi_{\mu_{Q^{-1} (1)} \cdots \mu_{Q^{-1}
(M)}},\; \cdot \;\rangle_M\; ,
\end{array}
 \end{equation}
 where $\mu_1, \cdots , \mu_M \leftrightarrow {\hat m}$
with $\mu_1 < \ldots < \mu_M$ is the correspondence defined by
(\ref{A1.13}). Now using the correspondence $\nu_1, \ldots \nu_N
\leftrightarrow {\hat n}$ with $\nu_1 < \ldots <\nu_N$ one finds that
\begin{equation}\label{3.15} (T ({\hat k}, \hat m) \otimes 1 \otimes
\ldots \otimes 1) \Psi^-_N (\hat n) = \Psi^-_M (\hat k) \otimes
{\chi(\hat n, \hat m)} ,
\end{equation} 
where
 \begin{align} \label{3.16}
 \chi(\hat n, \hat m) &= (N! M!)^{-\frac{1}{2}}\, \sum_{\substack{P \in
\mathcal S_N\\ Q \in \mathcal S_M}} \sigma^- (P) \sigma^-
(Q)\\\notag
&\phantom{\chi(\hat n, \hat m) = (N!M!)^{-\frac{1}{2}}}
\D{(\Pi^M_{j=1} \langle \phi_{\mu_{Q^{-1}(j)}} ,
\phi_{\nu_{P^{-1}(j)}} \rangle_1) \phi_{\nu_{P^{-1} (M+1)} \ldots
\nu_{P^{-1} (N)}}}.
\end{align}          
 Thus finally we obtain the relation
\begin{equation} \label{3.17} Z (\hat n', \hat n, \hat k, \hat m) =
\langle \Psi^-_N (\hat n'), \Psi^-_M (\hat k) \otimes \chi (\hat n,
\hat m) \rangle.
\end{equation}
{\bf 3.3.2:} In this subsection the following \textit
{proposition} is proved:
\begin{equation} \label{3.18} \chi (\hat n, \hat m) \not= 0 ,
\end{equation}
if and only if $\hat n - \hat m \in BZF$.

Firstly it is assumed that $\hat n - \hat m \notin BZF$. Then there is
a number $\alpha$ such that $m_\alpha = 1$ and $n_\alpha =
0$. Consequently, for each $Q \in {{\mathcal{S}}_M}$ there is an $r$
for which $Q^{-1}(r) = \alpha$ holds. But for each $P \in
{{\mathcal{S}}_N}$ the relation $P^{-1} (r) \not= \alpha$ is
true. Thus for each pair $P,Q$
\begin{equation} \label{3.19} \Pi^M_{j=1} \langle \phi_{\mu_{Q^{-1}
(j)}}, \phi_{\nu_{P^{-1}(j)}} {\rangle}_1 = 0
\end{equation}
so that also $\chi (\hat n, \hat m) = 0$.

Secondly let us assume that $\hat n - \hat m \in BZF$. Then for each
$\alpha$ with $m_\alpha = 1$ also $n_\alpha = 1$ holds. Consequently
one has to look for all pairs $Q,P$ such that
\begin{equation} \label{3.20} \Pi^M_{j=1} \langle \phi_{\mu_{Q^{-1}
(j)}}, \phi_{\nu_{P^{-1} (j)}} \rangle_1 = 1.
\end{equation}
For all other pairs $Q, P$ the product in (\ref{3.20})
is zero because $\langle \phi_\mu, \phi_\nu \rangle_1 = \delta_{\mu
\nu}$. Thus (\ref{3.20}) is equivalent to
\begin{equation} \label{3.21} \mu_{Q^{-1}(j)} = \nu_{P^{-1} (j)},\quad
j = 1, \cdots, M.
\end{equation}
In order to satisfy (\ref{3.21}), for a given $Q \in
{\mathcal{S}}_M$ the permutation $P \in {\mathcal{S}}_N$ must be such
that the $\mu_1, \cdots, \mu_M$, which by presumption occur in $\nu_1,
\cdots, \nu_M$, occupy the places $1, \cdots, M$ being ordered by
$Q$. All these pairs $Q, P$ can be explicitly indicated by the
following procedure.

Let $S \in {\mathcal{S}}_N$ be that permutation for which
\begin{equation} \label{3.22} (\nu_{S^{-1}(1)}, \cdots,
\nu_{S^{-1}(N)}) = (\mu_1, \cdots, \mu_M, \varrho_1, \cdots,
\varrho_{N-M}),
\end{equation}
where $\varrho_1, \cdots, \varrho_{N-M}$ are all those
$\nu_1, \cdots, \nu_N$ which are unequal $\mu_1, \cdots, \mu_M$, and
in addition let $\varrho_1 < \cdots < \varrho_{N-M}$. Then with the
help of (\ref{A1.7}) one finds
\begin{equation} \label{3.23}
\begin{array} {ll} \Psi^-_N (\hat n) &= \sigma^- (S) \sqrt{N!} S^-_N
U(S) \phi_{\nu_1 \cdots \nu_N} \\ [3ex] &= \sigma^- (S) \sqrt{N!}
S^-_N \phi_{\mu_1 \cdots \mu_M \varrho_1 \cdots \varrho_{N-M}}
\end{array}
\end{equation}
so that (\ref{3.16}) now reads
\begin{equation} \label{3.24}
\begin{array}{ll} \chi (\hat n, \hat m) = (N! M!)^{-\frac{1}{2}}
\sigma^- (S) \sum_{P,Q} \sigma^- (P) \sigma^- (Q)\quad \quad
\quad\\[3ex] \phantom{\chi (\hat n, \hat m) = N! M!}(\Pi^M_{j=1}
\langle \phi_{\mu_{Q^{-1}(j)}}, \phi_{\mu_{P^{-1} (j)}} \rangle)
\phi_{\varrho_{P^{-1}(M+1)-M} \cdots \varrho_{P^{-1}(N-M)}} \; .
\end{array}
\end{equation}
It follows from (\ref{3.24}) that only those pairs
$Q,P$ give nonzero summands for which $P$ has the form:
\begin{equation} \label{3.25} P = \left(Q, \begin{array}{cc} M + 1,
\cdots, N\\ M +1, \cdots, N \end{array} \right) \left(
\begin{array} {cc} 1, \cdots, M \\ 1, \cdots, M
\end{array}, R \right),
\end{equation}
where $R \in {\mathcal{S}}_{N-M}$ is an arbitrary
permutation. Hence, for a given $Q$ there are $(N-M)!$ permutations
$P$ of the form (\ref{3.25}) such that (\ref{3.21}) is satisfied.

Since $R$ acts on $(\nu_{S^{-1}(M+1)}, \cdots \nu_{S^{-1}(N)})=
(\varrho_1, \cdots, \varrho_{N-M})$ one finally obtains
\begin{equation} 
\begin{array}{ll} \label{3.26} \chi (\hat n, \hat m) &= (N!
M!)^{-\frac{1}{2}} \sigma^- (S) \sum_{RQ} \sigma^- (R) \sigma^- (Q)^2
\phi_{\varrho_{R^{-1} (1)} \cdots\varrho_{R^{-1}(N-M)}} \\ [3ex] &= 
{\binom NM}^{-\frac{1}{2}} \sigma^- (S) \Psi^-_{N-M} (\hat n - \hat m)
\end{array}
\end{equation} 
for all pairs $\hat n, \hat m$ with $\hat n - \hat m
\in BZF$.  Hence the proof of relation (\ref{3.18}) is complete, and
in addition the explicit form of $\chi (\hat n, \hat m)$ is
obtained.\\ 
{\bf 3.3.3:} Inserting (\ref{3.26}) into (\ref{3.15}) and
(\ref{3.13}) yields
\begin{equation} \label{3.27} Z (\hat n', \hat n, \hat k, \hat m) = 
{\binom NM}^{-\frac{1}{2}} \sigma^- (S) \langle \Psi^-_N (\hat n'),
S^-_N (\Psi^-_M (\hat k) \otimes \Psi_{N-M} (\hat n - \hat m)) \rangle
.
\end{equation} With the help of (\ref{A1.7}) and by the correspondence
$\hat k \leftrightarrow (\kappa_1, \cdots, \kappa_M)$ it follows that
\begin{equation} \label{3.28}
\begin{array} {ll} S^-_N (\Psi^-_M (\hat k) \otimes \Psi^-_{M-N} (\hat
n - \hat m)) = \\ [3ex] = (M! (N-M)!)^{-\frac{1}{2}} S^-_N
(\phi_{\kappa_1 \ldots, \kappa_M} \otimes \phi_{\varrho_1 \cdots
\varrho_{N-M}})\\ [3 ex] ={\binom NM}^{-\frac{1}{2}} \sigma^- (T)
\Psi^-_N (\hat n + \hat k - \hat m),
\end{array}
\end{equation}
where $T \in \mathcal{S}_N$ is the permutation which
lines up the sequence $(\kappa_1, \cdots, \kappa_M,\\ \varrho_1,
\cdots, \varrho_{N-M})$ in its natural order. Thus one obtains
\begin{equation} \label{3.29}
  \begin{array} {ll} Z (\hat n', \hat n, \hat k, \hat m) = \\ [3ex] =
    {\binom NM}^{-1} \sigma^- (S \cdot T) \langle \Psi^-_N (\hat
    n'),\Psi^-_N ( \hat n + \hat k-\hat m) \rangle\\ [3ex] =
    {\binom NM}^{-1} \sigma^- (S \cdot T) \delta (\hat n', \hat n + 
    \hat k - \hat m).
\end{array}
\end{equation} 
It follows that $Z (\hat n', \hat n, \hat k, \hat m )
\not= 0$ exactly if $\hat n - \hat m \in BZF, \hat n + \hat k - \hat m
\in BZF_N$ and $\hat n' = \hat n + \hat k - \hat m$.\\ 
{\bf 3.3.4:}
Since the permutations $S$ and $T$ are uniquely defined by the
sequences of indices $(\nu_1, \cdots, \nu_N)$, $(\mu_1, \cdots,
\mu_M)$ and $(\kappa_1, \ldots, \kappa_M)$ or equivalently by $\hat n,
\hat m$ and $\hat k$ it is obvious to define the function $C$ by
\begin{equation} \label{3.30}
\begin{array}{ll} C (\hat n, \hat k, \hat m) \;= \sigma^- (T \cdot S)
= \pm 1, \mathit{if} \; \hat n - \hat m \in BZF \;\mathit{and} \; \hat
n + \hat k - \hat m \in BZF_N\\ \quad\quad\quad\quad\quad\; =
0,\quad\quad\quad\quad\quad\quad \text{otherwise}.
\end{array}
\end{equation}
Now inserting (\ref{3.29}) together with (\ref{3.30})
into (\ref{3.12}) Formula (\ref{3.11}) is seen to hold, thus also
Lemma (3.5). With respect to the sums $\sum_{\hat k}$ and $\sum_{\hat
m}$ I refer to Remark 3.3 and Conclusion 3.6.

\subsection{An algorithm for $ C(\hat n, \hat k, \hat m)$}
{\bf 3.4.1:}
The question to be answered in this section reads: is there a finite
procedure for calculating $C (\hat n, \hat k, \hat m)$ if $\hat n,
\hat k, \hat m$ are given $bzf$. As in Section 3.3 the more general
case $2 \leq M < N$ is considered.

Since by definition $C (\hat n, \hat k, \hat m) = 0$ if the condition
\begin{equation} \label{3.31} \hat n - \hat m \in BZF\quad
\mathit{and} \quad \hat n + \hat k - \hat m \in BZF_N
\end{equation}
does not hold, only the case needs to be considered
that (\ref{3.31}) is true. Then
\begin{equation} \label{3.32} C(\hat n, \hat k, \hat m): = \sigma^- (T
\cdot S) = (-1)^{J(T)} (-1)^{J(S)},
\end{equation}
where $S$ is defined by (\ref{3.22}) and $T$ by
(\ref{3.28}). Moreover, $J(P)$ here means the number of inversions of
a permutation $P$ (cf. e.g. (\ref{A1.16})).\\ 
{\bf 3.4.2:} To begin
with, $J(S)$ is to be calculated. Let $\hat n$ be given. Then exactly
$N$ numbers $\nu_i, i=1, \cdots, N$ exist such that $n_{\nu_i} = 1$
and $\nu_1 < \cdots <\nu_N$. Hence $\hat n \leftrightarrow (\nu_1
\cdots, \nu_N)$. Likewise, if $\hat m$ is given, exactly $M$ numbers
$\mu_j, j=1, \cdots M$ exist such that $m_{\mu_j} = 1$ and $\mu_1 <
\cdots < \mu_M$.

Because of $\hat n - \hat m \in BZF$ for each $j \in \{1, \cdots M\}$
there is an $r_j$ such that
\begin{equation} \label{3.33} \mu_j = \nu_{r_j} \; \mathit{and} \; j
\leq r_j \;.
\end{equation}
The permutation $S$ is defined by (\ref{3.22}), i.e.
\begin{equation} \label{3.34} (\nu_{S^{-1}(1)}), \cdots,
\nu_{S^{-1}(N)}) = (\mu_1, \cdots, \mu_M, \varrho_1, \cdots,
\varrho_{N-M})
\end{equation}
and $\varrho_1 < \cdots < \varrho_{N-M}$. Then the
right-hand side of (\ref{3.34}) can be generated from $(\nu_1, \cdots,
\nu_N)$ by the following procedure.

First, $\mu_1 = \nu_{r_1}$ is positioned at the $r_1-th$ place in
$(\nu_1, \cdots, \nu_N)$. Therefore one needs $r_1 - 1$ inversions to
bring $\mu_1$ at the first place. Thereby the positions of $\mu_2,
\cdots, \mu_M$ in $(\nu_1, \cdots,\nu_N)$ are not changed.

Second, $\mu_2 = \nu_{r_2}$ is positioned at the ${r_2}^{th}$ place in
$(\nu_1 \cdots, \nu_N)$ so that one needs $r_2 - 2$ inversions to
bring $\mu_2$ at the second place. Again the positions of $\mu_3,
\cdots, \mu_M$ are not changed.

Thus, in order to bring $\mu_j = \nu_{r_j}$ to position $j$ one needs
$r_j - j$ inversions. Therefore the total number of inversions, which
realize the permutation $S$ in (\ref{3.34}), is given by
\begin{equation} \label{3.35} J(S) = \sum_{j=1}^M r_j - \frac{1}{2} M
(M+1).
\end{equation}
{\bf 3.4.3:} Now, $J(T)$ is to be determined. This task
is the following. Let $\hat n - \hat m$ and $\hat k$ be given. Then
$\hat n - \hat m \in BZF$ corresponds to the sequence of indices
$(\varrho_1, \cdots, \varrho_{N-M})$ and $\hat k$ to the sequence
$(\kappa_1, \cdots, \kappa_M,)$. Thus the sequence of indices
$(\kappa_1, \cdots, \kappa_M, \varrho_1 \cdots, \varrho_{N-M})$,
brought to its natural order by the permutation $T$ and denoted
$(\nu_1', \cdots, \nu'_N)$, corresponds to $\hat n + \hat k - \hat m
\in BZF$. Hence
\begin{equation} \label{3.36} (\nu'_{T(1)}), \cdots, \nu'_{T(N)}) =
(\kappa_1, \cdots, \kappa_M, \varrho_1, \cdots, \varrho_{N-M}).
\end{equation}
Therefore, because $\hat n + \hat k - \hat m$ is given,
also $(\nu'_1, \cdots, \nu'_N)$ is determined so that for each $j \in
\{1, \ldots, M\}$ the position $s_j$ of $\kappa_j$ in $(\nu'_1,
\cdots, \nu'_N)$ can be read off with the help of the relations
\begin{equation} \label{3.37} \kappa_j = \nu'_{s_j} \; \mbox{and}\; j
\leq s_j \;.
\end{equation}
Using the same arguments as in Section 3.4.2 one
obtains for $T^{-1}$ the result
\begin{equation} \label{3.38} J(T) = J(T^{-1}) = \sum_{j=1}^M s_j
-\frac{1}{2} M (M+1)\;.
\end{equation}
{\bf 3.4.4:} Finally, the algorithm for $C (\hat n,
\hat k\ , \hat m)$ can be formulated thus:\\ {\bf 1$^{st}$ step:} Take
$bzf\;\hat n, \hat k\ \text{and}\ \hat m$ which fulfil the equations
$\sum {n_\alpha} =N$, $\sum k_\alpha = \sum_\alpha m_\alpha = M$, and
test Condition (\ref{3.31}). If it is satisfied go to the next
step. If it is not, define $C (\hat n, \hat k, \hat m) = 0$, so that
the task has been done. \\ {\bf 2$^{nd}$ step:} Take $\hat n, \hat m$
and determine the corresponding sequences of indices $(\nu_1, \cdots,
\nu_N)$ and $(\mu_1, \cdots, \mu_M)$. Then from (\ref{3.33}) read off
the numbers $r_j, j=1, \cdots, M$, and calculate $J(S)$ with the help
of (\ref{3.35}).\\ {\bf 3$^{rd}$ step:} Take $\hat n + \hat k - \hat
m$ and $\hat k$, and determine the corresponding sequences $(\nu'_1,
\cdots, \nu'_N)$ and $(\kappa_1, \cdots, \kappa_M)$. Then from
(\ref{3.37}) read off the numbers $s_j, j=1, \cdots, M$ and calculate
$J(T)$ with the help of (\ref{3.38}).\\ {\bf 4$^{th}$ step:} Calculate
$C (\hat n, \hat k, \hat m)$ using (\ref{3.32}).\\ The coefficients $C
(\hat n, \hat k, \hat m)$ do not depend on the specific physical
system, for which they are used, rather they are completely
combinatorial. In other words, they result solely from the algebraic
structure imposed on the set $BZF$. Therefore they can be
computationally calculated once for all. A trivial special result is
the following.\\ If $\hat k = \hat m$, then $T^{-1} = S$ so that
\begin{equation} \label{3.39} C (\hat n, \hat m, \hat m) = 1.
\end{equation}

\subsection{The final form of $\langle\hat n' |H^-_N| \hat n \rangle$}
{\bf 3.5.1:} For the sake of simplicity in this section only the
special case $M = 2$ is considered. This does not entail any loss,
because the Hamiltonians we are interested in this paper are supposed
to have two-particle interactions. The starting point for this section
therefore is (\ref{3.7}). Moreover it is assumed that the matrix
elements $E(\hat k, \hat m)$ of the dummy Hamiltonian $H^-_{20}$ and
the coefficients $C (\hat n, \hat k, \hat m)$ are given.

Now the {\it problem} to be solved reads as follows. \textit{Let the
pair $\hat n', \hat n$ be given. Then determine those pairs $\hat k,
\hat m$ for which the summands in $\sum_{\hat k}$ and $\sum_{\hat m}$
do not vanish on general grounds.}

It is known from the previous considerations that for given $\hat n',
\hat n$ only those $\hat k$ and $\hat m$ in (\ref{3.7}) are relevant
which satisfy the condition
\begin{equation} \label{3.40} \hat n - \hat m \in BZF, \quad \hat n +
\hat k - \hat m \in BZF_N,\quad \hat n' = \hat n + \hat k - \hat m,
\end{equation}
hence also
\begin{equation} \label{3.41} \hat n' - \hat k \in BZF, \quad \hat n'
- \hat n = \hat k - \hat m .
\end{equation}
Consequently, the sums in (\ref{3.7}) have only  finitely many summands as already 
remarked in Conclusion 3.3.\\ 
{\bf 3.5.2:} In this section a disjoint
dissection of all possible pairs $\hat k, \hat m$ for given $\hat n',
\hat n$ will be defined. For this purpose it is useful to introduce
some new\\ {\bf Notation 3.8:} Let $\hat k, \hat m \in BZF_2$ so that
$\sum k_\alpha =\sum m_\beta = 2$ is satisfied. Then the sequence
$\hat d:= \hat k - \hat m$ is called differences sequence. The set of
all difference sequences is denoted $\mathcal{D}$.

Consequently there are only three types ${\mathcal{D}}_{\varrho},
\varrho = 0, 1, 2$ of possible $\hat d$ generated by $\hat d = \hat k
- \hat m:$\\ ${\mathcal{D}}_0$ contains only one element $\hat o: =
(0, \cdots, 0,\cdots)$, and $\hat o$ is generated by all $\hat k, \hat
m$ with $\hat k = \hat m$.\\ ${\mathcal{D}}_1$ contains $\hat d = (1,
-1, 0, \cdots)$ and all permutations of it. They are generated by all
$\hat k, \hat m$ such that for exactly one $\alpha$ the relation
$k_\alpha = m_\alpha = 1$ holds.\\ ${\mathcal{D}}_2$ contains $\hat d
= (1, 1, -1,-1, 0, \cdots)$ and all permutations of it. They are
generated by all $\hat k, \hat m$ for which no $\alpha$ exists such
that $k_\alpha = m_\alpha = 1$.\\ It follows immediately that the sets
${\mathcal{D}}_0,{\mathcal{D}}_1, {\mathcal{D}}_2, $ are disjoint and
that
\begin{equation} \label{3.42} {\mathcal{D}} = {\mathcal{D}}_0, \cup
{\mathcal{D}}_1 \cup {\mathcal{D}}_2.
\end{equation}
Many results of the next sections and chapters are
based on the following\\ 
{\bf {Proposition 3.9:}} For each pair $\hat
n', \hat n \in BZF_N$ there is a set
\begin{equation} \label{3.43} \{\hat d_1, \ldots, \hat d_L \}
\subset{\mathcal{D}}_1
\end{equation}
such that
\begin{equation} \label{3.44} \hat n' = \hat n + \hat d_1 + \ldots
\hat d_L.
\end{equation}
The number $L$ is uniquely determined by $\hat n'$ and
$\hat n$, but the difference sequences $\hat d_j, j = 1, \ldots, L$
are not.\\ 
{\bf{Proof:}} From the pair $\hat n', \hat n$ one forms the
matrix
\begin{equation} \label{3.45} X: = \left( \begin{array}{c} \hat n\\
\;\hat n'
\end{array} \right) = \left(
\begin{array}{c} n_1, n_2, \cdots \\ n'_1, n'_2, \cdots
\end{array} \right).
\end{equation} 
For each column of $X$ the following alternative holds:
\begin{equation} \label{3.46} \binom{n_\varrho}{n'_\varrho} = \binom
  00\ \text{or}\  {\binom 11}\  \text{or}\ \binom 01\ \text{or} \binom 10.
\end{equation} 
Because $\hat n$ and $\hat n'$ both contain $N$ numbers
1, there are equally many columns $\binom 01$ and $\binom10$.
Let $L$ by the number of each of the two kinds. Moreover let
$\varrho_j$ and $\sigma_j$, $j=1, \cdots, L$ be numberings of the
indices of these columns such that
\begin{equation} \label{3.47} \binom{n_{\varrho_j}}{n'_{\varrho_j}}
= \binom 01 \quad \text{and} \quad \binom{n_{\sigma_j}}{n'_{\sigma_j}}
=\binom 10.
\end{equation}  
Then for each pair of indices $\varrho_j, \sigma_j, j=
1, \cdots, L$ define $\hat d_j = (d_{j1}, d_{j2}, \cdots )$ by $d_{j
\varrho_j} = 1, d_{j \sigma_j} = -1$ and $d_{j\alpha} = 0, \alpha
\not= \varrho_j, \sigma_j$.\\ It follows from (\ref{3.47}) that
\begin{equation} \label{3.48} \hat n + \hat d_j = (\cdots, n_\alpha,
\cdots, n'_{\varrho_j}, \cdots, n'_{\sigma_j}, \cdots, n_\beta,
\cdots)\;,
\end{equation}
if $\varrho_j < \sigma_j$, and analogously, if
$\sigma_j < \varrho_j$.\\ Since $j \not= i$ implies $\varrho_j \not=
\varrho_i, \sigma_j \not= \sigma_i$ the addition of $\hat d_i$ to
$\hat n + \hat d_j$ can be carried through without altering
$n'_{\varrho_j}$ and $n'_{\sigma_j} $ in (\ref{3.48}). Thus, finally
one ends up with (\ref{3.44}) so that the proposition is proved.\\
Three immediate {\it consequences} are useful later on.

\begin{enumerate}
\item [1.)] For each pair $i, j$ with $i \not= j$ the relation $\hat
d_i + \hat d_j \in {\mathcal{D}}_2$ holds.
\item [2.)] $\hat n' - \hat n \in {\mathcal {D}}_\varrho$, exactly if
$L = \varrho,\; \varrho = 0, 1, 2$.
\item [3.)] $\hat n' - \hat n \notin {\mathcal D}$ exactly if $L \geq
3$.
\end{enumerate} 
{\bf 3.5.3:} Using the results of the previous
sections the problem formulated in 3.5.1 now can be solved by giving a
disjoint classification of the matrix elements defined by
(\ref{3.7}). According to (\ref{3.42}) four cases have to be taken
into account.\\ {\bf 1$^{st}$ case:} $\hat n' - \hat n \notin
{\mathcal{ D}}$. It follows immediately from (\ref{3.7}) that
\begin{equation} \label{3.49} \langle \hat n' | H^-_N | \hat n\rangle
= 0.
\end{equation} 
{\bf 2$^{nd}$ case:} $\hat n' - \hat n \in {\mathcal
D}_0$, i.e. $\hat n' = \hat n$. Then $\langle \hat n | H^-_N | \hat
n\rangle$ is unequal zero only, if also $\hat k = \hat m$. Hence the
double sum $\sum_{\hat k} \sum_{\hat m}$ reduces to a single sum
$\sum_{\hat m}$. This sum runs over all $\hat m$ for which $\hat n -
\hat m \in BZF$ holds. Since $\hat n$ contains $N$ numbers 1 there are
exactly $\binom N 2$ different sequences $\hat m$ such that this
condition is satisfied. Because of $C (\hat n, \hat m, \hat m) = 1$,
define
\begin{equation} \label{3.50} {\mathcal{E}} (\hat n, \hat o) =
\sum_{\hat m} E (\hat m, \hat m)
\end{equation}
for all $\hat m$ with $\hat n - \hat m \in BZF$. Thus
finally
\begin{equation} \label{3.51} \langle \hat n | H^-_N| \hat n \rangle =
{\mathcal{ E}} (\hat n, \hat o)\;.
\end{equation}
{\bf 3$^{rd}$ case:} $\hat n' - \hat n = \hat d_1 \in
{\mathcal{ D}}_1$. Hence $\hat n + \hat d_1 \in BZF$. Then $\langle
\hat n' |H^-_N| \hat n \rangle$ is unequal zero only, if $\hat k =
\hat m + \hat d_1$.  Therefore the double sum $\sum_{\hat k}
\sum_{\hat m}$ again reduces to a single sum $\sum_{\hat m}$ which
runs over all $\hat m$ so that $\hat n - \hat m \in BZF$ and $\hat m +
\hat d_1 \in BZF$. These $\hat m$ can be characterized as follows.

Let $\hat d_1$ be given by $d_{1\kappa} = 1, d_{1\mu} = -1$ and $
d_{1\beta} = 0, \beta \not= \kappa, \mu$. Hence $n_\kappa = 0$ and
$n_\mu = 1$. Then $\hat m + \hat d_1 \in BZF$ if and only if $m_\kappa
= 0$ and $m_\mu = 1$. In order to satisfy the condition $\hat n - \hat
m \in BZF$ it is necessary and sufficient that $n_\mu = 1$ and that
there is an $\alpha \not= \kappa, \mu$, for which $n_\alpha = 1$ and
$m_\alpha = 1$ holds.  Since $n_\kappa = 0$ and $n_\mu = 1$ there are
$N-1$ numbers $\alpha \not= \kappa,\mu$ for which $n_\alpha = 1$ so
that $\sum_{\hat m}$ runs over all $\hat m$ for which $m_\alpha =
m_\mu = 1$. Now define
\begin{equation} \label{3.52} {\mathcal{E}} (\hat n, \hat d_1) =
\sum_{\hat m} C(\hat n, \hat m + \hat d_1, \hat m) E (\hat m + \hat
d_1, \hat m).
\end{equation}
 Then
\begin{equation} \label{3.53} \langle \hat n' | H^-_N| \hat{n} \rangle
= {\mathcal{E}} (\hat{n}, \hat{d_1}),
\end{equation}
if $\hat n' = \hat{n} + \hat{d_1},\: \hat{d_1} \in
{\mathcal{D}}_1$\\ {\bf 4$^{th}$ case:} $\hat n' - \hat n = \hat d_2
\in {\mathcal{ D}}_2$. Let $\hat d_2$ be defined by $d_{2 \kappa} =
d_{2 \lambda} = 1, d_{2\mu} = d_{2\nu} = -1$. Then $\hat k$ and $\hat
m$ with $\hat k - \hat m = \hat d_2$ are uniquely determined by
$k_\kappa = k_\lambda = 1$ and $m_\mu = m_\nu =1$. Hence the
$\sum_{\hat h} \sum_{\hat m}$ reduces to a single term. Now define
\begin{equation} \label{3.54} {\mathcal{E}} (\hat n, \hat d_2) =
C(\hat{n}, \hat m + \hat d_2, \hat m) E (\hat m + \hat d_2, \hat m)\;,
\end{equation}
where $\hat m$ is determined by $ m_\mu = m_\nu =
1$. Then, if $\hat{n}'-\hat{n} = d_2 \in {\mathcal{D}}_2$
\begin{equation} \label{3.55} \langle \hat n' | H^-_N | \hat n \rangle
= {\mathcal{E}} (\hat{n}, \hat{d_2}).
\end{equation} 
{\bf 3.5.4:} Summing up one obtains\\
{\bf Proposition 3.10:} 1.) The matrix elements of the fermionic
Hamiltonian $H^-_N$ defined in Assumption 3.1 (Section 3.1.1.) are
given by
\begin{equation} \label{3.56}
\begin{array}{lll} \langle \hat n' |H^-_N| \hat{n} \rangle &=
{\mathcal{E}} (\hat{n}, \hat n' - \hat{n}), &\mbox{if} \quad \hat n' -
\hat n \in \mathcal{D} \\ &= 0 , &\mbox{if} \quad \hat n' - \hat n
\notin {\mathcal{D}}
\end{array}
\end{equation}
and
\begin{equation} \label{3.57} {\mathcal{E}} (\hat n', \hat n - \hat
n') = \bar{\mathcal{E}} (\hat n, \hat n' - \hat n).
\end{equation}
2.) Let be given $\hat n \in BZF$ and $\hat d \in
{\mathcal{D}}$ so that $\hat n + \hat d \in BZF.$ Then it follows from
Formula (\ref{3.7}) that
\begin{equation} \label{3.58} {\mathcal{E}} (\hat n, \hat d) =
\sum_{\hat m} C (\hat n, \hat m + \hat d, \hat m) E (\hat m + \hat d,
\hat m)\; ,
\end{equation}
where the sum runs over all $\hat m$ for which $\hat n
- \hat m \in BZF$ and $\hat m + \hat d =: \hat k \in BZF_2$. Hence,
the matrix elements $ \langle \hat n' |H^-_N| \hat{n} \rangle $ are
determined solely by the matrix elements $E (\hat k , \hat m)$ of $
H^-_{20} $, for which $ \hat k -\hat m = \hat d = \hat n' -\hat n
$. The sum in (\ref{3.58}) is finite.

For bosons a result holds, which is formally equal to (\ref{3.7}),
(\ref{3.56}) and (\ref{3.58}), but the terms $C, E, {\mathcal{E}}$ are
defined differently.

\section{Heuristic Considerations} \setcounter{equation}{0} 
{\bf 4.1:}
The result of Chapter 3, summarized in Formula (\ref{3.56}), is
completely formal up to now. This is due to two unsolved questions
connected with it. They read as follows.\\ \textit {First, can
Assumption 3.1 be verified? More concretely, is it possible to find a
one-particle Hilbert space ${\mathcal{H}}^1$ such that the dummy
Hamiltonian $H^-_{20}$ is defined on ${\mathcal{H}}^2_-$ and is
bounded, and such that in addition the spectrum of $H^-_{20}$ contains
the discrete eigenvalues of the dummy Hamiltonian $\bar{H}^-_{20}$,
which is primarily defined ?} (Cf. e.g. (\ref{A2.1}), (\ref{A2.4}),
(\ref{A2.6}).)\\ \textit{Second, is Formula (\ref{3.56}) of any
advantage for the spectral problem of $H^-_N$?}\\ 
{\bf 4.2:} To begin
with, let us look for an answer to the first question. As is already described 
in Subsection 2.1, any
investigation of the Hamiltonian $\bar H_N$ of an $N$-particle system
starts with a more or less informal specification of the external
fields acting upon the particles and of their
interactions. Customarily this is done using the position-spin
representation. Then $\bar H_N$ is of the form (\ref{1.1}) or, what is
the same, (\ref{2.1}). It is densely defined in a Hilbert space $\bar
{\mathcal{H}}^N$. Likewise the corresponding dummy Hamiltonian
$\bar{H}^-_{20}$ can be immediately written down as is shown for two
examples in appendix A2. It is defined in a dense linear submanifold
of $\bar {\mathcal{H}}^2$.

In order to verify Assumption 3.1 the space $\bar{\mathcal{H}}^1$ has
to be properly restricted to a subspace ${\mathcal{H}}^1$. Such a
restriction in turn can be carried through by finding a proper
orthonormal system ${\mathcal{O}}_1$ in $\bar{ \mathcal{H}}^1$ so that
${\mathcal{H}}^1 = \mathit{span}\; {\mathcal{O}}_1$. Having in mind
the physical meaning of ${\mathcal{H}}^1$ suggests   
taking the Hartree-Fock procedure for $\bar H^-_{20}$ to
determine ${\mathcal{O}}_1$. (For the details cf. Appendix A.3.) This is because this procedure is based
on the Ritz variational principle which guarantees optimal
approximation. Disregarding the fact that the Hartree-Fock procedure
generally is an infinite task, let us assume that it is completely
carried through for the dummy Hamiltonian $\bar H^-_{20}$ in
$\bar{\mathcal{H}}^2_-$. Thus one has obtained ${\mathcal{O}}_1$ and an orthonormal system
${\mathcal{O}}_2 \subset \bar{\mathcal{H}}^2_-$ of vectors
\begin{equation} \label{4.1} \Psi^-_{\kappa \lambda} =
\frac{1}{\sqrt{2}} (\phi_{\kappa} \otimes \phi_\lambda - \phi_\lambda
\otimes \phi_\kappa),\; \kappa < \lambda .
\end{equation}\\
By definition, ${\mathcal{O}}_1$ is an ONB of ${\mathcal{H}}^1$, therefore  ${\mathcal{O}}_2$ is an ONB of ${\mathcal{H}}_2^-$.
The corresponding energy levels $E_{\kappa
\lambda} = E( \hat m, \hat m)$ for $\kappa, \lambda \leftrightarrow
\hat m$ approximate the discrete eigenvalues of $\bar H^-_{20}$
outside its continuous spectrum.
Since the Hamiltonians considered in this paper are supposed to have a
bounded discrete spectrum, the set of the energy levels ${E_{\kappa
\lambda}}$ is also bounded.

Then the restriction $H^-_{20}$ of $\bar{H}^-_{20}$ to the space
${\mathcal{H}}^2_-$ is bounded because its spectrum is approximated by
the set $\{E_{\kappa \lambda}: \kappa < \lambda\}$ and its
eigenvectors by the set ${\mathcal{O}}_2$. As usual $H^-_{20}$ can be
defined on the whole space ${\mathcal{H}}^2_-$ using its matrix
representation with respect to the ONB \:${\mathcal{O}}_2$.\\
In most cases of practical application the complete Hartree-Fock
procedure cannot be achieved, because it is infinite. Therefore one
has to content oneself with a finite section of this procedure. But 
also such a finite procedure can be complicated.\\ 
Thus other methods were invented which
are equivalent to the Hartree-Fock procedure or approximate it.

{\bf{4.3:}} Let us now come to the second question. From (\ref{3.56})
one draws immediately a simple consequence.\\ 
{\bf Proposition 4.1:}
The matrix of the Hamiltonian $H^-_N = \Omega^-_N(H^-_{20})$ is
diagonal if the matrix of the dummy Hamiltonian $H^-_{20}$ is
diagonal.\

Unfortunately this result cannot be used to obtain the exact discrete eigenvalues of  a realistic N-particle system, because the exact eigenvectors of a dummy Hamiltonian $H^-_{20}$ containing interaction are not elements of any ONB $\mathcal{B}^-_2$. But, if one contents with a Hartree-Fock approximation of the eigenvalues of $H^-_{20}$, Proposition 4.1 delivers a Hartree-Fock-like approximation for the eigenvalues of $H^-_N$. 

Thus, if one wants to obtain better approximations, one has to solve the following     \textit{problem}. Since the results of Chapter 3 are valid for arbitrary orthonormal bases $\mathcal {B}_1 \subset \mathcal {H}^1$, one firstly has to choose such an ONB and one has to calculate the matrix elements $E (\hat k, \hat m)$ of 
$\mathcal{H}^-_{20 }$ for the ONB $\mathcal{B}^-_2$. The second part of the problem then is the question, wether the choise of $\mathcal{B }_1$ is helpful for a reasonable approximation of $\mathcal {H}^-_{20}$ and also of $\mathcal{H}^-_N$.\\ 
A heuristic idea to cope with this problem is the following. Choose the ONBs $\mathcal{B}_1$, resp.  ${\mathcal{B}}^-_2$ such that the matrix $E$ of $H^-_{20}$ with
respect to ${\mathcal{B}}^-_2$, i.e. the matrix defined by the
elements $E (\hat k, \hat m)$ is "as diagonal as possible". This means
the elements of ${\mathcal{B}}^-_2$ should approximate the
eigenvectors of $H^-_{20}$ optimally. Hence we end up again with the
Hartree-Fock method, an equivalent of it or an approximation.
Therefore $\mathcal{B}_1 = \mathcal{O}_1$ and ${\mathcal{B}}^-_2 = {\mathcal{O}}_2$ with $\mathcal{O}_1$ and ${\mathcal{O}}_2$ being defined in Section 4.2 and in Appendix 3.

In order to get further insight into the general structure of the
matrix $E$ I will give some purely heuristic arguments, which are
based on physical intuition. For this purpose let the Hartree-Fock
energy levels $E_{\mu \lambda}$ be numbered such that $E_{\mu \nu}
\leq E_{\mu \lambda}$ if $\mu < \nu < \lambda$. Then, if $\lambda$ is
large enough, i.e. if it exceeds a certain value $\bar \alpha$ one
expects that the particle having state $\lambda$ is "almost" free, so
that the interaction between the two particles having the states $\mu$
and $\lambda$ is "almost" zero. Thus the two particles with states
$\mu$ and $\lambda$ are "almost" free, if one particle of this pair is
"almost" free. This implies, that the vector $\Psi^-_{\mu \lambda} \in
{\mathcal{H}}^2_-$ is "almost" an eigenvector of $H^-_{20}$. Now let
$\mu, \lambda$ correspond to a sequence of occupation numbers $\hat
m$. Then $E(\hat m, \hat m)$ is "almost" an eigenvalue of $H^-_{20}$
for the eigenvector $\Psi^-_{\kappa \lambda } = \Psi^-_{2} (\hat m)$,
so that $E (\hat k, \hat m)$ is "almost" equal to $E (\hat m, \hat m)
\delta (\hat k, \hat m)$. Consequently $E (\hat k, \hat m)$ is
"small", i.e. it is "almost" zero, if $\hat k \not= \hat m$.

Now let us suppose that the term "small" has been concretized. Then
the above considerations can be summarized in the following {\it
assumption}.

{\it There is a natural number $\bar \alpha$ such that $E (\hat k,
\hat m)$ is small for each pair $\hat k, \hat m$ with $\hat k \not=
\hat m$, for which a $k_\sigma=1, \sigma > \bar \alpha$ exists or an
$m_\varrho = 1, \varrho > \bar \alpha$.}

This suggests truncating the matrix $E$ by substituting zeros for its
small elements. Then intuitively one conjectures that the truncated
matrix leads to an approximate solution of the spectral problem of
$H^-_N$ which is the goal of this paper.  The precise meaning of the
conjecture will be given in Section 5.1. \\ {\bf 4.4:} The first step
is defining the truncated dummy Hamiltonian $\hat H^-_{20}$ and
drawing some consequences. The operator $\hat H^-_{20}$ is determined
by its matrix $\hat E$, which is given by the elements:
\begin{equation}\label{4.3}
\begin{array}{l} \hat E (\hat k, \hat m)= 0,\; \mbox{if} \; \hat k
\not= \hat m \;\mbox{and if} \; \sigma, \varrho \; \mbox{exist such
that} \\ \phantom{\hat E (\hat k, \hat m)= 0,\;\; } k_\sigma = 1,\;
\sigma > \bar{ \alpha}\;\; \mbox{or} \;\; m_\varrho = 1,\; \varrho >
\bar {\alpha}, \\ \hat E (\hat k, \hat m)= E (\hat k, \hat m), \;
\quad\mbox{otherwise} .
\end{array}
\end{equation} 
According to this definition $\hat E$ is a finite
nondiagonal matrix of order $\binom{\bar \alpha} 2$ with an infinite
diagonal tail of elements $E (\hat m, \hat m)$ where $\hat m$ contains
at least one $m_\varrho = 1,\; \varrho > \hat \alpha$.

Consequently, $\hat H^-_{20}$ is bounded so that $\hat H^-_{N} =
\Omega^-_N (\hat H^-_{20})$ is also bounded and has the form
(\ref{3.56}), but with $\hat {\mathcal{E}}(\hat n, \hat d)$ defined by
$\hat E (\hat k, \hat m)$ the same way as ${\mathcal{E}} (\hat n, \hat
d)$ is determined by $E (\hat k, \hat m)$. Then we obtain the
following \\ {\bf Proposition 4.2:} Let us consider $\langle\hat n'
|\hat H^-_N| \hat n \rangle$.\\ 1.) If $\hat n' - \hat n \notin
{\mathcal{D}}$, one concludes from (\ref{3.7}) that $\hat n' - \hat n
\not= \hat k - \hat m$ which implies (cf. (\ref{3.44}))
\begin{equation} \label{4.4} \langle\hat n' |\hat H^-_N| \hat n
\rangle = \langle \hat n' | H^-_N| \hat n\rangle = 0.
\end{equation}
2.) If $\hat n' - \hat n = \hat d \in {\mathcal{D}}$
and if there is $\lambda > \bar \alpha$ such that $d_\lambda \not= 0$,
it follows from $\hat d = \hat k - \hat m$ that $k_\lambda = 1$ or
$m_\lambda = 1$. Hence $\hat E (\hat k, \hat m)=0$, and consequently
\begin{equation} \label{4.5} \langle\hat n' |\hat H^-_N| \hat n
\rangle = 0.
\end{equation}
3.) Now let $\hat n' - \hat n = \hat d \in
{\mathcal{D}}$ and suppose that $d_\lambda = 0$, if $\lambda >\bar{
\alpha}$. Then if $\hat d \in {\mathcal{D}}_2$, the $bzf$ $\hat k$ and
$\hat m$ are uniquely determined and one obtains (cf. Formula
(\ref{3.54}))
\begin{equation} \label{4.6} \langle\hat n' |\hat H^-_N| \hat n\rangle
= \hat{\mathcal{E}} (\hat n, \hat d) = {\mathcal{E}} (\hat n, \hat d)
= \langle\hat n' | H^-_N| \hat n\rangle.
\end{equation}
If $\hat d \in {\mathcal{D}}_1$, things are more
complicated. One has to apply the general method described in Section
3.5.3 for several special cases. The matrix element\\$\langle\hat n'
|\hat H^-_N| \hat n\rangle$ can be zero or unequal zero, and it need
not be equal to $\langle\hat n' |H^-_N| \hat n\rangle$.\\ {\bf4.5:} By
these considerations the problem of determining the eigenvalues of
$H^-_N$ is transformed into the following two ones.\\ \textit{First,
do the eigenvalues of $\hat H^-_N$ approximate those of $H^-_N$?} \\
\textit{Second, can the eigenvalues of $\hat H^-_N$ be calculated, at
least partially?}\\ The first problem is treated in Chapter 5, the
second one in Chapter 6.

\section{Spectral Approximation} 
\setcounter{equation}{0} {\bf 5.1:}
The dummy operators $H^-_{20}$ and $\hat H^-_{20}$ are defined on
${\mathcal{H}}^2_-$ and are bounded so that $D: = H^-_{20} -
{\hat{H}}^-_{20}$ is also bounded and defined on
${\mathcal{H}}^2_-$. Moreover ${\hat{H}}^-_{20}$ and $D$ depend on the
fixed number $\bar \alpha$. In what follows these operators are
understood to be functions of a parameter $\alpha \in \mathbb{N}$ and
$\alpha \geq 2$ so that they are written $D(\alpha)$ and $\hat
H^-_{20} (\alpha)$, and for the matrix elements of $\hat E$ we write
$\hat E_\alpha (\hat k, \hat m)$. Then by definition
\begin{equation} \label{5.1} \langle \hat{k} |D (\alpha)| \hat{m}
\rangle = E (\hat{k}, \hat{m}) - \hat{E}_\alpha (\hat{k}, \hat{m}),\;
\alpha \geq 2.
\end{equation}
 Therefore
\begin{equation} \label{5.2}
\begin{array}{l} \langle \hat{k} |D (\alpha)| \hat{m} \rangle = 0,\;
\mbox{if} \; \hat{k} = \hat{m} \; \mbox{or} \; \hat k \leftrightarrow
\kappa_1, \kappa_2 \leq \alpha \; \mbox{and} \; \hat m \leftrightarrow
\mu_1, \mu_2 \leq \alpha, \\ [2ex]

\langle \hat{k} |D (\alpha)| \hat{m} \rangle = \langle \hat{k} |D (2)|
\hat{m} \rangle, \; 
\kappa_1 > \alpha\; \mbox{or} \;\kappa_2 > \alpha \; \\
\phantom{\langle \hat{k} |D (\alpha)| \hat{m} \rangle = \langle
\hat{k} |D (2)| \hat{m} \rangle, \mbox{if} \hat{k} \not= \hat{m}}
\quad \mbox{or} \;\mu_1 > \alpha \; \mbox{or} \; \mu_2 > \alpha.
\end{array}
\end{equation}
These properties have the following consequence. \\
{\bf Proposition 5.1:} The sequence $(D (\alpha) : \alpha \in
\mathbb{N}, \alpha \geq 2) $ converges strongly to $0$.\\ {\bf Proof.}
Let the projections $F_\alpha$ and $F'_\alpha$ be defined by
\begin{equation} \label{5.3} \displaystyle{F_\alpha =
\sum^\infty_{\kappa_1, \kappa_2 > \alpha } \Psi^-_{\kappa_1 \kappa_2}
\langle \Psi^-_{\kappa_1, \kappa_2}, \;\cdot\; \rangle }
\end{equation}
and $F'_\alpha = 1 - F_\alpha$\;. Then $F_\alpha$
converges strongly to $0$, if $\alpha \rightarrow \infty$, and
$F'_\alpha$ to $1$.  By a simple calculation using Formula (\ref{5.2})
one verifies that for each $f \in {\mathcal{H}}^2_-$:
\begin{equation} \label{5.4}
\begin{array}{ll} F_{\alpha} D (\alpha)f = F_\alpha D (2) f, \\ [3ex]
F'_\alpha D (\alpha)f = F'_\alpha D (2) F_\alpha f.
\end{array}
\end{equation}
Therefore
\begin{equation} \label{5.5} \| D(\alpha) f \| \leq || F_\alpha D(2) f
\| + \| F'_{\alpha} D (2) F_\alpha f \|.
\end{equation}
Because of
\begin{equation} \label{5.6} \| F'_\alpha D(2) F_\alpha f \| \leq \| D
(2) \| \| F_\alpha f \|
\end{equation}
and because $F_\alpha$ converges to $0$, the
proposition is seen to hold.

Now, in order to transfer the last result to $\hat H^-_N (\alpha) : =
\Omega^-_N (\hat H_{20} (\alpha))$ one needs the following theorem.\\
{\bf Proposition 5.2:} Let be given a sequence $(A_M (r): r \in
\mathbb{N})$ of bounded operators defined on $\mathcal{H^M}$. If $A_M
(r)$ converges strongly to $0_M$ for $r \rightarrow \infty$, then also
\begin{equation} \label{B1.33} s - \lim_{r - \infty} \Omega^-_N (A_M
(r)) = 0_N
\end{equation}
i.e. $\Omega^-_N (A_M (r))$ converges strongly.\\ {\bf
Proof:} 1.) Let $g \in {\mathcal H}^N$, then
$$g = \sum_{\kappa_1 \dots \kappa_N} b_{\kappa_1 \dots \kappa_N} \phi_{\kappa_1 \dots \kappa_N}.$$\\
If
$$ \chi_{\kappa_{M+1} \ldots \kappa_N} := 
\sum_{\kappa_1 \dots \kappa_M} b_{\kappa_1 \dots \kappa_N}
\phi_{\kappa_1 \dots \kappa_M}\;, $$\\ it follows that
$$ g = \sum_{{\kappa_{M+1}} \dots \kappa_N} \; \chi_{\kappa_{M+1} \dots \kappa_N} \; \otimes \phi_{\kappa_{M+1} \dots \kappa_N}$$\\
and
$$ \| g \|^2 = \sum_{\kappa_{M+1} \dots \kappa_N} \| \chi_{\kappa_{M+1}\dots  \kappa_N} \|^2.$$\
If $B$ is a bounded operator on ${\mathcal H}^M$, then
\begin{equation} \label{B1.34} \| (B \otimes 1 \otimes \dots \otimes
1) g \|^2 = \sum_{\kappa_{M+1}\ldots \kappa_N} \| B \chi_{\kappa_{M+1}
\dots \kappa_N} \|^2.
\end{equation}
2.) Now, let us consider the operators $A_M (r)$. By
supposition $A_M(r)$ converges strongly to $0_M $. Hence by the
principle of uniform boundedness (cf. e.g. \cite{Kato}, p. 150) there
is a number $K$ such that for all $r \in \mathbb{N}$:
$$ \| A_M (r) \| \leq K. $$\\
Thus, one obtains for all $r \in \mathbb{N}$:
\begin{equation} \label{B1.35} \sum_{\kappa_{M+1}\ldots \kappa_N} \|
A_M (r) \chi_{\kappa_{M+1} \ldots \kappa_N} \|^2 \leq \quad K^2
\sum_{\kappa_{M+1} \dots \kappa_N} \| \chi_{\kappa_{M+1} \ldots
\kappa_N} \|^2.
\end{equation}
Hence, by the criterion of {Weierstra\ss} the left
series converges uniformly with respect to the variable $r$. Therefore
the limit $r \rightarrow \infty$ can be interchanged with the sum so
that
\begin{equation} \label{B1.36}
\begin{array} {ll} \lim_{r \rightarrow \infty} \| (A_M (r) \otimes 1
\otimes \dots \otimes 1) g \|^2 \\ [4ex]= \sum_{\kappa_{M+1} \ldots
\kappa_N} \lim_{r \rightarrow \infty} \| A_M (r)
\chi_{\kappa_{M+1}\ldots \kappa_N} \|^2 &= 0\;.
\end{array}
\end{equation}
Thus, because for all $g \in {\mathcal H}^N_-$ the
relation
\begin{equation*}
\|\Omega^-_N (A_M (r)) g \| \leq {\binom NM} \| (A_M (r)\otimes 1
\otimes \dots) g \|
\end{equation*}
holds, Formula (\ref{B1.33}) is proved.

 Then, with the help of the Propositions 5.1 and 5.2 one obtains the
following consequences.\\ 
{\bf Proposition 5.3:} 1.) The sequence
($\hat H^-_N (\alpha): \alpha \in \mathbb{N}, \alpha \geq 2$)
converges strongly to $H^-_N$. In other words, the operators $\hat
H^-_N (\alpha)$ approximate $H^-_N$ (cf. \cite{Chat}, p. 228, Formula
\ref{5.1}). Hence all results concerning the approximation of one
operator by a strongly converging sequence of operators are valid for
$H^-_N$ and $\hat H^-_N (\alpha)$. (Cf. \cite{Chat}, Chapter 5.). Here
only two of these properties are sketched.\\ 2.) If $e$ is an isolated
eigenvalue of $H^-_N$, then there is a sequence ($e_\alpha: \alpha \in
\mathbb{N}$) such that $e_\alpha$ is an eigenvalue of $\hat H^-_N
(\alpha)$ and such that $e_\alpha \rightarrow e$. (Cf. \cite{Chat},
p. 239, Theorem 5.12).\\ 3.) Let $P$ be the spectral projection
belonging to an eigenvalue $e$ of $H^-_N$. Then an $\alpha_0$ exists
such that for each $\alpha > \alpha_0$ there is a spectral projection
$P_\alpha$ belonging to $\hat H^-_N (\alpha)$ and such that $P_\alpha$
converges strongly to $P$ for $\alpha \rightarrow
\infty$. (Cf. \cite{Chat}, p. 240, Theorem 5.13.)\\ 
{\bf 5.2:} Besides
the approximation of $H^-_N$ by $ \hat H^-_N(\alpha)$ sketched above
there are results, which are based on the norm of $D (\alpha)$,
i.e. on $\delta (\alpha):= \| D (\alpha) \|$. Using the Formulae
(\ref{A1.29}) and (\ref{A1.30}) one obtains the relation
\begin{equation} \label{5.7} \| H^-_N - \hat H^-_N (\alpha) \| =
\parallel \Omega^-_N (D (\alpha))\parallel \leq {\binom N2} \delta
(\alpha).
\end{equation}
Thus it follows that $\hat H^-_N (\alpha)$ converges in
norm to $H^-_N$ if
\begin{equation} \label{5.8} \lim_{\alpha \rightarrow \infty} \delta
(\alpha) = 0.
\end{equation}
Theorems concerning spectral approximation based on
convergence in norm can be found in \cite{Chat}, p. 291, Theorem 4.10,
p. 362, Theorem 5.10 and in \cite{Kato}, p. 249, Proposition 5.28.

\section{The operator $\hat H^-_N$ and its matrix}
\subsection{Preliminary remarks} \setcounter{equation}{0}

{\bf 6.1.1:} The aim of this chapter is proving \\ {\bf Proposition
6.1:} 1.) The operator $\hat{H}^-_N$ defined on ${\mathcal{H}}^N_-$ is
an orthogonal sum of operators defined on finite dimensional
(orthogonal) subspaces of ${\mathcal{H}}^N_-$.\\ 2.) Moreover, using
the matrix of ${\hat H}^-_N$ the matrices of the suboperators can be
determined explicitly.

If this proposition is verified, we have obtained a
block-diagonalization of ${\hat H}^-_N$. Thus a way is opened,
depending on the numbers $\bar \alpha$ and $N$, to calculate the
eigenvalues of $\hat H^-_N$ with the help of numerical methods. Purely
analytical solutions of the eigenvalue problem of $\hat H^-_N$ are
also possible, if $\bar \alpha = 2, 3, 4.$ But, in these cases one
cannot expect that $\hat H^-_N$ is a good approximation of a realistic
Hamiltonian $H^-_N$.\\ {\bf 6.1.2:} In order to verify Proposition 6.1
some further notation is used, which is provided by\\ {\bf Definition
6.2:} 1.) Let be given an $\hat n \in BZF_N$ and a natural number
$\alpha$, which for the moment is completely arbitrary. Then
\begin{equation} \label{6.1} (\hat n,\alpha) := (n_1, \cdots ,
n_\alpha) \quad \mbox{and} \quad (\alpha, \hat n): = (n_{\alpha+1},
n_{\alpha + 2}, \cdots).
\end{equation}
For the infinite second part of $\hat n$ also the
abbreviation $(\alpha, \hat n) = : \hat r$ is used.\\ 2.) Let
$\hat{r}$ be given. Then ${\mathcal{N}}_\beta (\hat r)$ denotes the
set of all $\hat n \in BZF_N$, for which $(\alpha, \hat n) = \hat r,
\; \sum^{\infty}_{\varrho = \alpha +1} n_\varrho = N- \beta$ and $0
\leq \beta \leq \mbox{min} \{\alpha, N \}$.  Hence the finite sequence
$(\hat n, \alpha)$ contains exactly $\beta$ numbers 1 and $\alpha -
\beta$ numbers $0$. Because $\beta$ is determined by $\hat r$, the
notation is a bit redundant, but it turns out to be useful.

Now it is supposed that $\alpha$ and $N$ are fixed numbers. Then,
Definition 6.2 yields the following\\ 
{\bf Consequence 6.3:} 1.) The
set ${\mathcal{N}}_\beta (\hat r),\; 0 \leq \beta \leq min \{\alpha,
N\}$ is finite, more precisely, card ${\mathcal{N}}_\beta (\hat r) =
{\binom \alpha \beta}$. Therefore the subspace ${\mathcal{H}}^N_-
(\hat r)$ of ${\mathcal{H}}^N_-$ spanned by the $\Psi^-_N (\hat n)$
for $\hat n \in {\mathcal{N}}_\beta (\hat r)$ has dimension $\binom\alpha
\beta$.\\ 
2.) The sets $\mathcal{N}_\beta (\hat r)$ and
$\mathcal{N}_{\beta'}(\hat r')$ are disjoint if $\hat r \not= \hat
r'$. Moreover $\beta = \beta'$, if and only if the sequences $\hat r ,
\hat r'$ contain the same number of elements 1. Then for each $ \hat n
\in \mathcal{N}_\beta (\hat r)$ there is an $ \hat n' \in
\mathcal{N}_\beta (\hat r')$ such that $(\hat n',\alpha) = ( \hat n,
\alpha)$.\\ 3.) Let $\mathcal{B}^-_N$ be the ONB of
${\mathcal{H}^N_-}$ defined by (\ref{A1.13}), and let $\Psi^-_N (\hat
n) \in \mathcal{B}^-_N$. Then there is exactly one $\beta$ so that
$\hat n \in \mathcal{N}_\beta (\alpha, \hat n)$.  Hence the sets
$\mathcal{N}_\beta (\hat r)$ with $\hat r = (n_{\alpha + 1}, n_{\alpha
+ 2}, \cdots)$ containing $N-\beta$ numbers 1 and $ 0 \leq \beta \leq
min \{\alpha, N\}$ form a complete disjoint dissection of the set of
all $\hat n \in BZF$ for which $\sum_\varrho n_\varrho =N$. \\ 4.)
From the above parts 2 and 3 one concludes that the spaces
$\mathcal{H}^N_- (\hat r)$ are orthogonal for different $\hat r$, and
that they span $\mathcal{H}^N_-$, i.e.
\begin{equation} \label{6.2} \mathcal{H}^N_- = \bigoplus_{\hat r}
\mathcal{H}^N_- (\hat r).
\end{equation}
Later on a restriction of the set $\mathcal{D}$ of all
difference sequences (cf. Notation 3.8) is needed.\\ {\bf Definition
6.4:} $\mathcal{D}_\alpha$ is the set of all $\hat d \in \mathcal{D}$,
for which $d_\varrho = 0$ if $\varrho > \alpha$. In addition let
$\mathcal{D}_{j \alpha}: = {\mathcal{D}}_j \cap {\mathcal{D}}_\alpha,
j =0, 1,2$.\\ Therefore the sets $\mathcal{D}_{j\alpha}$ are again
disjoint, and ${\mathcal{D}}_{0 \alpha} = {\mathcal{D}}_0 = \{ \hat
o\}$.\\ {\bf{6.1.3:}} Finally a lemma is proved which is basic for the
further considerations.\\ {\bf Proposition 6.5:} If $\hat n \in
\mathcal{N}_\beta (\hat r), 0 \leq \beta \leq \mathit{ min} \{\alpha,
N\} $ and if $\hat d \in \mathcal{D}_\alpha,\hat d \not= \hat o$, then
either $\hat n + \hat d \in \mathcal{N}_\beta (\hat r)$ or $\hat n +
\hat d \notin BZF$.\\ {\bf Proof:} The proof is complete if one can
show that $\hat n + \hat d \in \mathcal{N}_{\beta} (\hat r)$ is
equivalent to $\hat n + \hat d \in BZF$. First, if $\hat n + \hat d
\in \mathcal{N}_\beta (\hat r)$, then $\hat n + \hat d \in BZF$
holds. Second, it follows from $\hat n + \hat d \in BZF$, that the
(two or one) numbers 1 in $\hat d$ must be at positions where there
are $0$ in $\hat n$. Likewise, the (two or one) numbers -1 in $\hat d$
must be at positions, where numbers 1 are in $\hat n$. Because of
$\hat d \in \mathcal{D}_\alpha$, the sequence $(\hat n + \hat d,
\alpha)$ has the same quantity $\beta$ of numbers 1 as $(\hat n,
\alpha)$ has, and $\hat r:= (\alpha, \hat n) = (\alpha, \hat n + \hat
d)$ because $\hat d$ does not affect $\hat r$. Thus $\hat n + \hat d
\in \mathcal{N}_\beta (\hat r)$, so that the proof is complete.

\subsection{General properties of the matrix of $\hat H^-_N$}

{\bf 6.2.1:} In what follows the definitions and results of Section
6.1 are applied for the special choice $\alpha =\bar \alpha$ with a
properly chosen $\bar \alpha$. Moreover, the ONBs \;
${\mathcal{B}}^-_2$ and $\mathcal{B}^-_N$ are those which are defined
via the Hartree-Fock procedure in the Sections 4.2 and 4.3.\\ {\bf
6.2.2:} In this subsection the first part of Proposition 6.1 is
proved.  In order to do so, the matrix representation of $\hat H^-_N$
with respect to $\mathcal{B}^-_N$ is used. The {\it proof} is
complete, if one shows that for $\hat n \in \mathcal{N}_\beta (\hat
r)$ and $\hat n' \in \mathcal{N}_{\beta'} (\hat r')$ with $\hat r
\not= \hat r'$:
\begin{equation} \label{6.3} \langle \hat n' |\hat H^-_N| \hat
n\rangle = 0.
\end{equation}
There are three possibilities for the pair $\hat n',
n$.\\ 
If $\hat n' - \hat n \notin \mathcal{D}$, it follows from
Formula (\ref{4.4}) that (\ref{6.3}) holds.\\ 
If $\hat n' - \hat n \in
{\mathcal{D}} \diagdown {\mathcal{D}}_{\bar \alpha}$, by Formula
({\ref{4.5}}) it is seen that (\ref{6.3}) holds, too.\\ 
If $\hat n' -\hat n \in {\mathcal{D}}_{\bar \alpha}$, then $\hat n' = \hat n + \hat
d \in BZF$ , $\hat d \in \mathcal{D}_{\bar\alpha} $ and $\hat d \not= \hat
o$. Thus it follows from Proposition 6.5 that $\hat n' \in
{\mathcal{N}}_\beta (\hat r)$. This result contradicts the supposition
$\hat n' \in \mathcal{N}_{\beta'} (\hat r')$ and $\hat r' \not= \hat
r$. Therefore $\hat n' - \hat n \notin \mathcal{D}_{\bar \alpha}$ is
true.\\ Hence (\ref{6.3}) holds if $(\bar {\alpha}, \hat n') \not=
(\bar { \alpha}, \hat n)$.

If one denotes the restriction of $\hat H^-_N$ to the space
$\mathcal{H}^N_- (\hat r)$ by $\hat H^-_N (\hat r)$ one obtains
\begin{equation} \label{6.4} \hat H^-_N = \bigoplus_{\hat r} \hat
H^-_N (\hat r).
\end{equation}
Thus, part one of Proposition 6.1 has been proved. The proof of the
second part is postponed to Section 6.3.\\ {\bf 6.2.3:} In this
subsection therefor some preparatory work will be done.  Since the
dimension of $\mathcal{H}^N_- (\hat r)$ is $\binom{\bar\alpha}\beta$
with $0 \leq \beta \leq {\mathit{min}} \{\bar \alpha, N\}$, it can
become gigantic depending on $N, \bar \alpha$ and $\beta$. Therefore,
aiming at the diagonalization of $\hat H^-_N (\hat r)$ it is of vital
interest to know how many matrix elements $\langle\hat n' |\hat H^-_N
(\hat r)| \hat n\rangle$ vanish on principle grounds. This means, how
many matrix elements of $\hat H^-_N$ are zero for arbitrary dummy
Hamiltonians $\hat H^-_{20}$ respectively their matrices $\hat E$.
Then in addition further matrix elements of $\hat H_N^-$ can be zero
for special $\hat H^-_{20}$. But the last aspect will not be
considered in this paper.

Now, from Consequence 4.3 one immediately draws\\ {\bf Conclusion
6.6:} Let $\hat n', \hat n \in \mathcal{N}_\beta (\hat r)$ and let
$\hat n' - \hat n \notin \mathcal{D}_{\bar \alpha}$, then
\begin{equation} \label{6.5} \langle \hat n' | \hat H^-_N | \hat n
\rangle = 0.
\end{equation}
Since by Proposition 3.9 the relation
$$\hat n' = \hat n + \hat d_1 + \cdots + \hat d_L, \quad \hat d_j \in \mathcal{D}_1$$\\
holds, one has a simple criterion to decide whether $\hat n' - \hat n
\notin \mathcal{D}_{\bar \alpha}$ or not. Especially, if $L > 2$
Formula (\ref{6.5}) is true.\\ {\bf{6.2.4:}} Finally the nondiagonal
matrix elements with $\hat n' - \hat n = \hat d \in \mathcal{D}_{\bar
\alpha}, \hat d \not= \hat o$ are considered. For this purpose let us
introduce the following\\ {\bf Notation 6.7:} If $\hat n', \hat n \in
BZF_N$. Then $\hat n', \hat n$ are called $\mathcal{D}_{j \bar
\alpha}-$concatenated, $j = 0,1,2$, if there is a $\hat d \in
\mathcal{D}_{j \bar \alpha}$ so that $\hat n' = \hat n + \hat d$. The
$bzf \;\hat n', \hat n$ are simply called $\mathcal{D}_{\bar\alpha}-$
concatenated if they are ${\mathcal{D}}_{j \bar \alpha}-$ concatenated
for $j=0$ or 1 or 2.\\ With the help of this notation we arrive at the
\\ {\bf Result 6.8:} 1.) For each $\hat n \in \mathcal{N}_\beta (\hat
r)$, there are exactly

\begin{equation*}
  \tau_1 (\bar \alpha, \beta) := \beta (\bar \alpha - \beta)
\end{equation*}
$\mathcal{D}_{1 \bar \alpha}-$concatenated $\hat n' \in
\mathcal{N}_\beta (r)$. This is because each number 1 out of the
$\beta$ numbers 1 in $(\hat n,\bar \alpha)$ can be transposed at each
position of the $ \bar \alpha - \beta$ numbers 0 by a $\hat d \in
\mathcal{D}_{1 \bar \alpha}$.\\ 2.) For each $\hat n \in
\mathcal{N}_\beta (\hat r)$ there are exactly
\begin{equation*}
\tau_2 (\bar \alpha, \beta) := {\binom\beta 2} {\bar \alpha -
  \binom\beta 2}
\end{equation*}
$\mathcal{D}_{2 \bar \alpha}$-concatenated $\hat n' \in
\mathcal{N}_\beta (\hat r)$, where ${\binom\varrho 2} = 0$ if
$\varrho = 0,1$. This holds because each pair of numbers 1 out of the
$ \beta$ numbers 1 in $(\hat n,\bar \alpha)$ can be brought at the
position of each pair out of the $\bar \alpha - \beta$ numbers 0 by a
$\hat d \in \mathcal{D}_{2 \bar \alpha}$.\\ 3.) $\hat n', \hat n \in
\mathcal{N}_\beta (\hat r)$ are $\mathcal{D}_{0 \bar \alpha}-$
concatenated exactly if $\hat n' = \hat n$.\\ 4.) For each $\hat n \in
\mathcal{N}_\beta (\hat r)$ there are exactly
\begin{equation} \label{6.6} \tau (\bar \alpha, \beta):= \tau_2 (\bar
\alpha, \beta) + \tau_1 (\bar \alpha, \beta) = {\binom\beta 2} {\bar
\alpha - \binom\beta 2} + \beta (\bar \alpha - \beta)
\end{equation}
$\mathcal{D}_{\bar\alpha}\text{-concatenated}\ \hat n'
\not= \hat n$.\\ 5.) Finally, let us consider the matrix of $\hat
H^-_N (\hat r)$ with the elements $\langle\hat n' |\hat H^-_N| \hat
n\rangle$, $\hat n', \hat n \in \mathcal{N}_\beta (\hat{r})$. Then, in
the $\hat n'$-row there are exactly $\tau (\bar \alpha, \beta)$
nondiagonal elements which can be unequal zero, and similarly for
$\hat n$-columns. Consequently, the number $Z (\bar \alpha, \beta)$ of
zero nondiagonal elements in each $\hat n'$-row or $\hat n$-column is
\begin{equation} \label{6.7} Z (\bar \alpha, \beta) = \binom{\bar \alpha}\beta - \tau (\bar \alpha, \beta) - 1.
\end{equation}
According to Conclusion 6.6\; $Z (\bar \alpha, \beta)$
is the number of $bzf \;\hat n$ for which $\hat n' - \hat n \notin
\mathcal{D}_{\bar \alpha}$ in any $\hat n'$-row, and likewise for the
$\hat n$-columns.

\subsection{ The matrices of the operators $\hat H^-_N (\hat r)$} In
this section the second part of Proposition 6.1 will be proved. This
runs as follows.\\ {\bf 6.3.1:} $\beta = \bar \alpha \leq N$. The
number $z$ of elements $\hat n \in \mathcal{N}_{\bar \alpha} (\hat r)$
is $\binom{\bar \alpha}{\bar \alpha} = 1$, and the element $\hat n$
has the form
\begin{equation} \label{6.8} \hat n = (1, \cdots, 1, n_{\bar \alpha +
1}, \cdots).
\end{equation}
Consequently the matrix of $\hat H^-_N (\hat r)$ is of
order one and its element is
\begin{equation} \label{6.9} \langle\hat n |\hat H^-_N| \hat n \rangle
= \hat{\mathcal{E}} (\hat n, \hat o) = \mathcal{E} (\hat n, \hat o).
\end{equation}
{\bf 6.3.2:} $\beta = \bar \alpha - 1 \leq N.$ 1.) The
number $z$ of elements $\hat n \in \mathcal{N}_{\bar \alpha - 1} (\hat
r)$ is $\binom{\bar \alpha}{\bar\alpha - 1} = \bar \alpha$, and the
$(\hat n, \bar \alpha)$ for $\hat n \in \mathcal{N}_{\bar \alpha-1}
(\hat r)$ contain only one 0 and $\bar \alpha - 1$ numbers 1. \\ 2.)
The $\hat n \in N_{\bar \alpha - 1} (\hat r)$ are numbered by $ \hat n
=: \hat n_\kappa$ if $0$ is at position $\bar \alpha - \kappa$ in
$(\hat n, \bar \alpha)$, and $\kappa = 0, \cdots, \bar \alpha - 1$.\\
3.) Any two elements $\hat n', \hat n \in \mathcal{N}_{\bar \alpha -1}
(\hat r)$ with $\hat n' \not= \hat n$ are $\mathcal{D}_{1 \bar
\alpha}$-concatenated. This is because the number of $\hat n \in
\mathcal{N}_{\bar \alpha - 1}(\hat r)$, which are $\mathcal{D}_{\bar
\alpha}$-concatenated with $\hat n' \not= \hat n$, according to
(\ref{6.6}) is
\begin{equation} \label{6.10} \tau (\bar \alpha, \bar \alpha - 1) =
\tau_1 (\bar \alpha, \bar \alpha -1) = \bar \alpha - 1 = z - 1.
\end{equation}
4.) The matrix of $\hat H^-_N (\hat r)$ in the present
case is a \;$z \times z= \bar \alpha \times \bar \alpha$ matrix, which
has the elements
\begin{equation} \label{6.11} \langle\kappa' |\hat H^-_N| \kappa
\rangle: = \langle\hat n_{\kappa'} |\hat H^-_N| \hat n_\kappa\rangle =
\hat{\mathcal{E}} (\hat n_\kappa, \hat n_{\kappa'} - \hat n_{\kappa}
).
\end{equation}
{\bf 6.3.3:} $\beta = \bar \alpha - 2 \leq N$. 1.) The
number $z$ of elements $\hat n \in \mathcal{N}_{\bar \alpha - 2} (\hat
r)$ is $\binom{\bar \alpha}{\bar \alpha - 2} = \frac{1}{2} \bar
\alpha (\bar \alpha -1)$, and the ($\hat n, \bar \alpha$) for $\hat n
\in \mathcal{N}_{\bar \alpha - 2} (\hat r)$ contain two numbers 0 and
$\bar \alpha - 2$ numbers 1.\\ 2.) The $\hat n \in \mathcal{N}_{\bar
\alpha-2} (\hat r)$ are numbered by $\hat n =: \hat n_{\kappa
\lambda}, \kappa < \lambda$ if the two $0$ are at the positions $\bar
{\alpha}-\lambda $ and $\bar \alpha - \kappa,\; \;0 \leq \kappa <
\lambda \leq \bar \alpha - 1$.\\ 3.) Any two elements $\hat n', \hat n
\in \mathcal{N}_{\bar \alpha - 2} (\hat r)$ with $\hat n' \not= \hat
n$ are $\mathcal{D}_{\bar \alpha}$-concatenated. This is because the
number of $\hat n \in \mathcal{N}_{\bar \alpha - 2}(\hat r)$, which
are $\mathcal{D}_{\bar \alpha}$-concatenated with $\hat n' \not= \hat
n$, according to (\ref{6.6}) is
\begin{equation} \label{6.12} \tau (\bar \alpha, \bar \alpha - 2) =
\binom{\bar \alpha}{\bar \alpha - 2} - 1 = z - 1.
\end{equation}
4.) The matrix of $\hat H^-_N (\hat r)$ then is a\; $z
\times z$ matrix with $z = \frac{1}{2} \bar \alpha (\bar \alpha-1)$,
which has the elements
\begin{equation} \label{6.13} \langle\kappa', \lambda' |\hat H^-_N|
\kappa, \lambda\rangle : = \langle\hat n_{\kappa' \lambda'} |\hat
H^-_N| \hat n_{\kappa \lambda}\rangle = \hat{\mathcal{E}} (\hat
n_{\kappa \lambda}, \hat n_{\kappa' \lambda'} - \hat n_{\kappa
\lambda}).
\end{equation}
{\bf 6.3.4:} $\beta = 2 < N$.  1.) The number $z$ of
elements $\hat n \in \mathcal{N}_2 (\hat r)$ is $\binom{\bar \alpha}2 
= \frac{1}{2} \bar \alpha (\bar \alpha - 1)$, and the $(\hat n,
\bar\alpha)$ for $\hat n \in \mathcal{N}_2 (\hat r)$ contain two
numbers 1 and $\bar \alpha - 2$ numbers 0.\\ 2.) The $\hat n \in
\mathcal{N}_2 (\hat r)$ are numbered by $\hat n =: \hat n_{\kappa
\lambda}, \kappa < \lambda$, if the two 1 are at positions $\kappa$
and $\lambda$, $1 \leq \kappa < \lambda \leq \bar \alpha$\\ 3.) Any
two $\hat n', \hat n \in \mathcal{N}_2 (\hat r)$ are
$\mathcal{D}_{\bar \alpha}$-concatenated. This follow via the same
argument as in 6.3.3.\\ 4.) Like in 6.3.3 one obtains the $z \times z$
matrix of $\hat H^-_N (\hat r)$. It has the elements
\begin{equation} \label{6.14} \langle\kappa' \lambda' |\hat H^-_N|
\kappa \lambda \rangle := \langle\hat n_{\kappa' \lambda'} |\hat
H^-_N| \hat n_{\kappa \lambda}\rangle = \hat{\mathcal{E}} (\hat
n_{\kappa \lambda}, \hat n_{\kappa' \lambda'}- \hat n_{\kappa
\lambda}).
\end{equation}
{\bf 6.3.5:} $\beta = 1, N>2$.\; 1.) The number $z$ of
elements $\hat n \in \mathcal{N}_1 (\hat r)$ is ${\bar \alpha \choose
1} = \bar\alpha$, and the $(\hat n, \bar \alpha)$ for $\hat n \in
\mathcal{N}_1 (\hat r)$ contain one number 1 and $\bar \alpha - 1$
numbers 0.\\ 2.) The $\hat n \in \mathcal{N}_1 (\hat r)$ are numbered
by $\hat n =: \hat n_{\kappa}$, if $1$ is at position $\kappa$, $1
\leq \kappa \leq \bar \alpha$.\\ 3.) Any two $\hat n', \hat n \in
\mathcal{N}_1 (\hat r)$ are $\mathcal{D}_{1 \bar
\alpha}$-concatenated. The argument is the same as in 6.3.2.\\ 4.)
Also as in 6.3.2 the matrix $ \hat H^-_N (\hat r)$ is obtained. It is
a $z \times z = \bar \alpha \times \bar \alpha$ matrix having the
elements
\begin{equation} \label{6.15} \langle\kappa' |\hat H^-_N|
\kappa\rangle:= \langle\hat n_{\kappa'} |\hat H^-_N| \hat
n_\kappa\rangle = \hat{\mathcal{E}} (\hat n_\kappa, \hat n_\kappa' -
\hat n_\kappa).
\end{equation}
{\bf 6.3.6:} $\beta = 0, N >2 $. The number $z$ of
elements $\hat n \in \mathcal{N}_0 (\hat r)$ is ${\bar \alpha \choose
0} = 1$, and the element $\hat n$ has the form
\begin{equation} \label{6.16} \hat n= (0, \cdots, 0, n_{\bar \alpha
+1}, \cdots ).
\end{equation}
Consequently the matrix of $ \hat H^-_N (\hat r)$ is of
order 1 and its element is
\begin{equation} \label{6.17} \langle\hat n |\hat H^-_N| \hat n\rangle
= {\hat {\mathcal{E}}} (\hat n, \hat o) = \mathcal{E} (\hat n, \hat
o).
\end{equation}
{\bf 6.3.7:} $2 < \beta < \bar \alpha - 2,\; \beta \leq
N$.\; 1.) The number $z$ of elements $\hat n \in \mathcal{N}_\beta\
(\hat r)$ is $\bar \alpha \choose \beta$. It is larger than the
numbers $z$ in the previous cases. Each element $\hat n \in
\mathcal{N}_\beta (\hat r)$ contains in $(\hat n, \alpha)$ at least
three numbers 1 and three numbers 0.\\ 2.) For each $\beta$ not all
pairs $\hat n', n' \in \mathcal{N}_\beta (\hat r)$ are
$\mathcal{D}_{\bar\alpha}$-concatenated. To prove this proposition it
suffices to give an example. Thus, let
\begin{equation} \label{6.18} \hat n' = (1,1,1, \cdots, 0,0,0, n_{\bar
\alpha+1}, \cdots),\; \hat n = (0,0,0, \cdots, 1,1,1, n_{\bar \alpha +
1} \cdots),
\end{equation}
and let $\hat d_j \in \mathcal{D}, j = 1,2,3$ be
defined by $d_{j \varrho} = \delta_{j \varrho} - \delta_{\bar \alpha +
j - 3, \varrho},\; \varrho \in \mathbb{N}$. Then
\begin{equation} \label{6.19} \hat n' = \hat n + \hat d_1 + \hat d_2 +
\hat d_3
\end{equation}
so that $\hat n' - \hat n \notin \mathcal{D}_{\bar
\alpha}$. The factual number of non concatinated elements can be
calculated from $Z(\bar \alpha, \beta)$ as defined by Formula
(\ref{6.7}).\\ 3.) The elements $\hat n \in \mathcal{N}_\beta (\hat
r)$ are numbered by $\hat n = \hat n_\kappa, \kappa = 1, \cdots, z$
arbitrarily.  Then the matrix elements of $ \hat H^-_N (\hat r)$ in
the present case are
\begin{equation} \label{6.20}
\begin{array} {lll} \langle\kappa' |\hat H^-_N| \kappa \rangle: &=
\langle \hat n_{\kappa'} |\hat H^-_N| \hat n_\kappa\rangle \\ &=
\mathcal{\hat E} (\hat n_\kappa, \hat n_{\kappa'} - \hat n_\kappa) &,
\hat n_{\kappa'} - \hat n_\kappa \in \mathcal{D}_{\bar \alpha} \\ &= 0
&, \hat n_{\kappa'} - \hat n_\kappa \notin \mathcal{D}_{\bar\alpha}\;.
\end{array}
\end{equation}
{\bf 6.3.8:} Besides the above properties of the
matrices of $\hat H^-_N (\hat r)$ the following result is of practical
relevance.\\ {\bf Proposition 6.9:} Let the sequences $\hat r$ and
$\hat r'$ have the same number $N-\beta$ of elements $1$. If $ \hat
n_1 ,\hat n_2 \in \mathcal{N}_\beta\ (\hat r)$ and $\hat n_1 \neq \hat
n_2$, there are $\hat n'_1 ,\hat n'_2 \in \mathcal{N}_\beta\ (\hat
r'), \hat n'_1 \neq \hat n'_2$ such that
\begin{equation} \label{6.21} \langle \hat n'_1 |\hat H^-_N (\hat r')|
\hat n'_2\rangle = \langle \hat n_1 |\hat H^-_N (\hat r)| \hat
n_2\rangle, \\
\end {equation} and vice versa. Thus, the matrices of $\hat H^-_N
(\hat r')$ and $\hat H^-_N (\hat r')$ have the same nondiagonal
elements.\\ {\bf Proof:} For given $ \hat n_1 ,\hat n_2$ the bzf $\hat
n'_1 ,\hat n'_2$ are chosen according to Consequence 6.3 as follows:
$(\hat n'_j, \alpha)=(\hat n _j, \alpha), j=1,2.$ Thus, $\hat n'_2 -
\hat n'_1 = \hat n_2 - \hat n_1$.\\ If $\hat n_2 - \hat n_1 \not\in
\mathcal{D}_{\bar\alpha}$, it follows from the proof in Subsection
6.2.2. that Formula (\ref{6.21}) holds, because both sides are zero.\\
Now let us assume that $ \hat d := \hat n_2 - \hat n_1 \in
\mathcal{D}_{\bar\alpha}$. Then applying Formula (\ref{3.58}) yields
\begin{equation} \label{6.22} \langle \hat n_1 |\hat H^-_N (\hat r)|
\hat n_2\rangle = \langle \hat n_1 |\hat H^-_N | \hat n_2\rangle =
\sum_{\hat m} C (\hat n_2, \hat m + \hat d, \hat m) E (\hat m + \hat
d, \hat m),
\end {equation} where the sum runs over all $\hat m$ with $\hat n_2
-\hat m \in BZF$ and $\hat k := \hat m + \hat d \in BZF_2 $. Because
$\hat E(\hat m + \hat d, \hat m) = 0$, if $k_\lambda = 1, \lambda >
\bar \alpha$ or $ m_\kappa = 1, \kappa > \bar \alpha $, in Formula
(\ref{6.22}) only such components $k_\rho, m_\sigma $ are relevant,
for which $\rho, \sigma \leq \bar \alpha.$ Therefore, if the condition
$ \hat n_2 - \hat m \in BZF$ is satisfied, then also $\hat n'_2 - \hat
m \in BZF $ holds. The other condition is also satisfied, because
$\hat d = \hat n'_2 - \hat n'_1$.  Finally, the inversions, which
determine $ C(\hat n, \hat m + \hat d, \hat m)$, only refer to the
elements of $(\hat n'_2, \alpha)=(\hat n _2, \alpha)$. Thus one
obtains
\begin{equation} \label{6.23} C(\hat n, \hat m + \hat d, \hat m) =
C(\hat n', \hat m + \hat d, \hat m).
\end{equation}
Hence the last term in (\ref{6.22}) is equal to
\begin{equation} \label{6.24} \sum_{\hat m} C (\hat n'_2, \hat m +
\hat d, \hat m) E (\hat m + \hat d, \hat m), = \langle \hat n'_1 |\hat
H^-_N | \hat n'_2\rangle = \langle \hat n'_1 |\hat H^-_N (\hat r')|
\hat n'_2\rangle.
\end{equation}
The proof of the inverse runs the same way.
\subsection {Conclusion} The decomposition of the matrix of $\hat
H^-_N$ into orthogonal finite matrices as described in the Sections
6.2 and 6.3 now allows, depending on $N$ and $\bar \alpha$, to
determine parts of the spectrum of $\hat H^-_N$. Thus, it is only a
question of the capacity of the computers available, which parts of
the spectrum one can calculate, and it is a question of physical
relevance, which parts one wants to calculate. Intuitively, the case
$\beta = \bar \alpha$ is a Hartree-Fock approximation. This suggests
that better approximations are achieved if $\bar \alpha$ is greater
than $N$, because then the case $\beta = \bar \alpha$ cannot occur.

\section{Final Remarks}

\subsection{Summary of the results} \setcounter{equation}{0}

{\bf 7.1.1:} The way of approaching the eigenvalue problem of the
Hamiltonian $H^-_N$ presented in the Chapters 3 to 6 may at the first
glance seem complicated.  Therefore it is useful to realize the simple
kernel of the procedure. I will present it in the form of a work
program which comprises six steps.\\ 
$\mathbf 1^{st}$ {\bf step:} As a
starting point one formulates the Hamiltonian $\bar H_N$ to be
considered. This is usually done making use of the position-spin
representation, i.e. $\bar H_N$ is an operator in the Hilbert space
$\bar{ \mathcal{H}}^N = \bigotimes^N (L^2(\mathbb{R}^{3}) \otimes
\mathcal{S}^1)$. The most general form of $\bar {H}_N$ for charged
particles is due to Breit. It can be found in the literature, e.g. in
\cite{Ludwig} p. 247.\\ 
{\bf 2$^{nd}$ step:} According to Formula
(\ref{2.4}) one shapes the dummy Hamiltonian $\bar H^-_{20}: =\bar
H^-_2 (\gamma_ 0), \gamma^{-1}_0 = N-1$, belonging to $\bar H_N$. Then
one determines the orhonormal system ${\mathcal{O}}_1$ via the Hartree-Fock
procedure for $\bar H^-_{20}$ (Cf. Appendix A.3.), an equivalent method or an
approximation. Eventually, the operators $\bar H^-_N$ and $\bar
H^-_{20}$ are restricted to the spaces $ \mathcal{H}^N_-,
\mathcal{H}^2_-$ with $\mathcal{H}^1= \mathit{span}\ {\mathcal{O}}_1$
and are denoted $H^-_N$ and $H^-_{20}$.\\ 
$\mathbf{3}^{rd}$ {\bf step:} In order to determine the matrix $E$ of $H^-_{20}$  one has to choose an ONB $\mathcal{B}_1 $ of $\mathcal{H}^1$. As explained in Section 4.3, the best choise is $\mathcal{B}_1 = \mathcal{O}_1.$ Then, using the ONB $\mathcal{B}^-_2 \subset \mathcal{H}^2_-$ the matrix elements  $E (\hat k, \hat m)$ are calculated (cf. e.g. (\ref{A1.10}), (\ref{A1.13})).\\ 
{\bf 4$^{th}$ step:} From the matrix $E$ one obtains the truncated matrix $\hat{E}$
by replacing the "small" elements $E (\hat k, \hat m), \hat k \not=
\hat m$ of $E$ by zeros as described in Section 4.4. The matrix $\hat
E$ depends on a number $\bar \alpha$ and determines the operator
$\hat H^-_{20}$. \\ 
{\bf 5$^{th}$ step:} The matrix elements
$\hat{\mathcal{E}}(\hat n, \hat d) $ of $\hat H^-_N = \Omega^-_N (\hat
H^-_{20})$ are calculated from the matrix elements $\hat E (\hat k,
\hat m)$ the same way as the $\mathcal{E} (\hat n, \hat d)$ are
calculated from $E (\hat k, \hat m)$ in Section 3.5. In this
connection the general results of the Sections 6.1 and 6.2 are
useful.\\
{\bf 6$^{th}$ step:} One determines the orthogonal
submatrices of the matrix of $\hat H^-_N$ according to Section
6.3 and diagonalizes them as many as possible numerically or
analytically.\\ Then one can try to obtain error estimates applying
suitable results of the theory of spectral approximation.\\ {\bf
7.1.2:} As mentioned at the end of Section 6.4, the lowest energy
levels of $\hat{H}^-_N$ are not simply Hartree-Fock-like approximations of
the true values, if $N$ is smaller than $\bar \alpha$. Thus in this
case the method presented here reproduces also those results, which
are obtained by other methods like density functional theory (DFT) or
configuration interaction method (CI). Summing up, it is intuitively
clear that the approximation of $H^-_N$ by $\hat H^-_N$ is the better
the smaller the number $N$ and the larger the parameter $\bar \alpha$,
which in turn is limited by the capacity of computers. My colleague
Arno Schindlmayr is preparing an application of the proposed method.

\subsection{Finite procedures}

{\bf 7.2.1:} The program described in Section 7.1.1 is a {\it work}\:
program in a strict sense only if all its steps could be carried
through in finite time. Thus, the critical points are found in those
steps which contain infinite tasks. The first and decisive one is the
determination of the infinite ONB \;$\mathcal{O}_1$ in the second
step. For one has to expect that in most cases ${\mathcal{O}}_1$ can
neither analytically nor numerically be calculated completely. Hence,
what can be performed is the determination of a finite part
$\mathcal{O}_{f1}$ of $\mathcal{O}_1$, i.e. its elements up to a
number $R$.

However, if only $\mathcal{O}_{f1}$ is available, the method described
here does not brake down. Rather the work program formulated in
Section 7.1.1 can also be carried through for $\mathcal{O}_{f1}$
instead of $\mathcal{O}_1$. The only question is, which of the
obtained results are of physical interest.

In order to get an answer let us use the following obvious {\it
notation}:
$$ \mathcal{B}_{f1} = \mathcal{O}_{f1},\mathcal{H}^1_f  , \mathcal{H}^2_{f-}, \mathcal{B}^-_{f2},
\mathcal{B}^-_{fN}, \mathcal{H}^N_{f-}, H^-_{f20}, \hat H^-_{f20},
H^-_{fN}, \hat H^-_{fN}, \hat E_f, \hat{\mathcal{E}}_f .$$ \\
Moreover, as already introduced above, $R$ is the number of elements
of $\mathcal{B}_{f1} $.

Now, because the reduced work program depends on the three parameters
$N, R$ and $\bar \alpha$, the above question can be answered as
follows.\\ 1.) If $N > R$, there are no vectors unequal zero in
$\mathcal{H}^N_{f-}$. Therefore this case has to be excluded.\\ 2.) If
$N = R$, the Hilbert space $\mathcal{H}^N_{f-}$ is $1$-dimensional, so
that the $N$-particle Hamiltonian $\hat H^-_{fN} = H^-_{fN}$ has one
eigenvalue of Hartree-Fock type. This result is of minor interest.

Thus, in order to get better results one needs an $R$ which is
"sufficiently" larger than $N$.\\ 3.) Let $\bar \alpha \leq N <
R$. Then, according to the classification described in Section 6.3,
for each $\beta$ satisfying \;$max \{0, \bar \alpha + N - R\} \leq
\beta \leq \bar \alpha$ \; at least one submatrix of the matrix of
$\hat H^-_{fN}$ exists, which in principle can be diagonalized
numerically.\\ 4.) Let $N < \bar \alpha < R$. Then, as in point 3, for
each $\beta$ with $max \{0, \bar \alpha + N - R\} \leq \beta \leq N$
there is again at least one submatrix of the matrix of $\hat
H^-_{fN}$, which in principle can be diagonalized numerically.\\ 5.)
If $N < \bar \alpha = R$, then $\beta = N$. Therefore, there is only
one submatrix of the matrix of $\hat H^-_{fN}$, which is identical
with the matrix of $\hat H^-_{fN}$. Moreover $\hat H^-_{fN} =
H^-_{fN}$. Hence, this case is optimal, but it possibly can not be
treated numerically because $R$ is too large.

\textit{The result of the above considerations now reads: only the
cases 3., 4. and 5. can be of physical interest.} \\ {\bf 7.2.2:}
Thus, the question arises, \textit{are} they. In other words, what can
be said about the spectrum of $\hat H^-_N$ by studying $\hat
H^-_{fN}$. The answer is given by\\ {\bf Proposition 7.1}: The
spectrum of $\hat H^-_{fN}$ is contained in the spectrum of $\hat
H^-_N$. Thus, the finite work program does not change the eigenvalues
of $\hat H^-_N$, rather it delivers only a subset of them.\\ The {\bf
proof} runs as follows. It suffices to show that the matrices of $\hat
H^-_{fN}$ and $\hat H^-_N$ with respect to the ONB\;
$\mathcal{B}^-_{fN}$ are identical.

Let $ {\mathcal{M}}_{fL}$ be the set of all $\hat l \in BZF_L$ such
that $l_\varrho = 0$, if $\varrho >R$. Then $\hat n \in
\mathcal{M}_{fN}$ exactly if $\Psi^-_N (\hat n) \in
\mathcal{B}^-_{fN}$. Now let $\hat k, \hat m \in BZF_2$ and $\hat n',
\hat n \in \mathcal{M}_{fN}$. If $\hat n - \hat m \in BZF$ and $\hat
n' - \hat k \in BZF$, then $\hat k, \hat m \in
{\mathcal{M}}_{f2}$. This implies
\begin{equation} \label{7.1} \hat E_f (\hat k, \hat m) = \hat E (\hat
k, \hat m) .\\
\end{equation}
In addition let $\mathcal{D}_f$ be the set of all $\hat
d:= \hat k - \hat m,\; \hat k, \hat m \in \mathcal{M}_{f2}$.  Thus
$d_\varrho = 0$, if $\varrho >R$, for each $\hat d \in \mathcal{D}_f$.

By definition, $\hat{\mathcal{E}}_f (\hat n, \hat d),$ where $\hat n
\in \mathcal{M}_{fN}$ and $\hat d \in \mathcal{D}_f$, is constructed
from $\hat E_f (\hat k, \hat m)$ via Formula (\ref{3.58}) like
$\hat{\mathcal{E}} (\hat n, \hat d)$ from $\hat E (\hat k, \hat m)$
(or $\mathcal{E} (\hat n, \hat d)$ from $E (\hat k, \hat
m))$. Therefore, by Formula (\ref{7.1}) one obtains
\begin{equation} \label{7.2} \hat{\mathcal{E}}_f (\hat n, \hat d) =
\hat{\mathcal{E}} (\hat n, \hat d)
\end{equation}
so that by use of (\ref{3.56})
\begin{equation} \label{7.3} \langle\hat n' | \hat H^-_{fN}| \hat
n\rangle = \langle\hat n' |\hat H^-_N| \hat n\rangle.
\end{equation}
Finally, if $\hat n \in \mathcal{M}_{fN}$ and $\hat n
\in \mathcal{N}_\beta (\hat r),\; \hat r = (\bar \alpha, \hat n)$,
then $\mathcal{N}_\beta (\hat r) \subset \mathcal{M}_{fN}$. Therefore,
corresponding submatrices of $\hat H^-_{fN}$ and $\hat H^-_N$ have the
same shape. Hence, they are identical.\\ This result guarantees the
practical applicability of the finite work program.

\setcounter{section}{0} \renewcommand\thesection{\Alph{section}}
\numberwithin{equation}{subsection}
\section{Appendix}
\subsection{Glossary} 

\textbf{A.1.1}\label{A.1.1}
The formalism briefly described in this section was mainly developed
by Cook \cite{Cook} and by Schroeck \cite{Schroeck}. The purpose of
this appendix is fixing notation and formulating some few results,
which are used throughout the paper.\\ The starting point is an axiom
of $QM$ that reads: Let $\mathcal{H}^1$ be the Hilbert space of a
system containing only one particle of a certain kind. Then the
Hilbert space of a system containing $N$ particles of the same kind is
the symmetric or the antisymmetric subspace of the $N-$fold tensor
product $\mathcal{H}^N: = \bigotimes^N \mathcal{H}^1$. Likewise the
Hilbert spaces of systems composed of different kinds of particles are
subspaces of appropriate tensor products of one-particle Hilbert
spaces.

In what follows the tensor product $\otimes$ of Hilbert spaces is
understood to be a complete space. But the noncomplete tensor product
of linear manifolds is a noncomplete linear manifold. This product is
denoted $\underline{\otimes}$.

The inner product in $\mathcal{H}^N$, denoted $\langle\;\cdot,
\;\cdot\;\rangle$ or $\langle\;\cdot, \cdot\;\rangle_N$, is defined as
usual by the inner product $\langle\;\cdot, \cdot\;\rangle_1$ in
$\mathcal{H}^1$ in the following way. If $f = f_1 \otimes \cdots
\otimes f_N$ and $ g = g_1 \otimes \cdots \otimes g_N$, then
\begin{equation} \label{A1.1} \langle f, g \rangle_N = \langle f_1,
g_1 \rangle_1 \cdots \langle f_N, g_N \rangle_1 \;.
\end{equation}
By linear and continuous extension $\langle\cdot,
\cdot\rangle_N$ is defined on $\mathcal{H}^N$.

The tensor structure of the $N$-particle Hilbert spaces implies the
following \\ 
{\bf Proposition A.1.1:} Let $\mathcal{B}_1 :=
\{\phi_\lambda : \lambda \in \mathbb{N}\}$ be an orthonormal basis
(ONB) in $\mathcal{H}^1$. Then
\begin{equation}\label{A1.2} {\mathcal{B}}_N := \{\phi_{\lambda_1}
\otimes \cdots \otimes \phi_{\lambda_N}: \lambda_j \in \mathbb{N},
j=1, \ldots, N \}\\
\end{equation}
is an ONB in $\mathcal{H}^N$.\\ Throughout this paper
the abbreviation is used:
\begin{equation} \label{A1.3} \phi_{\lambda_1 \cdots \lambda_N} =
\phi_{\lambda_1} \otimes \cdots \otimes \phi_{\lambda_N} .
\end{equation}
{\bf A.1.2:} Let ${\mathcal S}_N$ be the symmetric
group, and let $P \in {\mathcal S}_N$. Then the operator $U (P)$ of
the exchange of particles is \textit{defined} by
\begin{equation} \label{A1.4} U (P) \phi_{\kappa_1 \cdots \kappa_N} =
\phi_{\kappa_{P^{-1}(1)} \cdots \kappa_{P^{-1}(N)}}.
\end{equation}
and by continuous linear extension.\\ The operator $U
(P)$ has the following properties.\\ {\bf Proposition A.1.2:} 1.) $U
(P)$ is invariant under a change of the ONB.\\ 
2.) $U(P)$ is defined
on $\mathcal{H}^N$ and is unitary, i.e. $U (P) U^\star (P) =
1$. Moreover
\begin{equation} \label{A1.5} U^\star (P) = U(P^{-1}),\; U(PQ) = U (P)
U (Q).
\end{equation}
With the help of $ U(P), P \in {\mathcal S}_N$ the
symmetrizer and the antisymmetrizer are {\it defined} by
\begin{equation} \label{A1.6} S^\pm_N = \frac{1}{N!} \sum_{P \in
{\mathcal S}_N} \sigma^\pm (P) U (P)
\end{equation}
with $\sigma^+ (P) = 1$ and $\sigma^- (P) =
(-1)^{J(P)}$, where $J(P)$ is either the number of inversions of $P$
or equivalently the number of transpositions forming $P$.

Some useful properties of the operators $S^\pm_N$ are summarized in
the next {\bf Proposition A.1.3:} $S^\pm_N$ are projections defined on
${\mathcal{H}}^N$. Moreover
\begin{equation} \label{A1.7}
\begin{array}{ll} U (P) S^\pm_N = S^\pm_N U(P) = \sigma^\pm (P)
S^\pm_N, \\ [3ex] S^\pm_{M+K} (S^\pm_M \phi_{\mu_1 \ldots, \mu_M}
\otimes S^\pm_K \phi_{\kappa_1 \ldots \kappa_K}) = S^\pm_{M+K}
(\phi_{\mu_1 \ldots \mu_M} \otimes \phi_{\kappa_1 \ldots \kappa_K}).
\end{array}
\end{equation}
Then the physically relevant subspaces of
${\mathcal{H}}^N$ are ${\mathcal{H}}^N_\pm = S^\pm_N [
{\mathcal{H}}^N]$ where $+$ stands for bosons and $-$ for
fermions. Thus
\begin{equation} \label{A1.8} {\mathcal{H}}^N = {\mathcal{H}}^N_+
\oplus {\mathcal{H}}^N_- \oplus {\mathcal{H}}^N_r ,
\end{equation}
 where $\oplus$ is the orthogonal sum as usual.

A special role in this paper play some orthonormal bases of
${\mathcal{H}}^N_\pm$, which are defined by \\ {\bf Proposition A.1.4:}
1.) The set ${\mathcal{B}}^+_N$ of all vectors
\begin{equation} \label{A1.9} \D{\Psi^+_{\kappa_1 \ldots \kappa_N} :=
\frac{\sqrt{N!}}{\sqrt{\Pi_j n_{\kappa_j}}!} S^+_N \phi_{\kappa_1
\ldots \kappa_N}}
\end{equation}
with $\kappa_1 \leq \ldots \leq \kappa_N$ and
$n_{\kappa_j} = \sum^N_{\alpha = 1} \delta_{\kappa_j\kappa_\alpha}$ is
an ONB in ${\mathcal{H}}^N_+$. Moreover $\sum_j n_{\kappa_j} = N$.\\
2.) The set ${\mathcal{B}}^-_N$ of all vectors
\begin{equation} \label{A1.10} \Psi^-_{\kappa_1 \ldots \kappa_N} :=
\sqrt{N!} S^-_N \phi_{\kappa_1 \ldots \kappa_N}
\end{equation}
with $\kappa_1 < \ldots < \kappa_N$ is an ONB in
${\mathcal{H}}^N_-$.\\ {\bf A.1.3:} For the problems to be treated in
this paper notation (\ref{A1.9}) and (\ref{A1.10}) is not optimal, it can be
improved by introducing the sequences of occupation numbers by the
following\\ {\bf Definition A.1.5:} 1.) Let be given a sequence of
indices $\kappa_1, \cdots, \kappa_N$ as in (\ref{A1.9}) or in
(\ref{A1.10}), where $\kappa_j \in \mathbb{N}, j = 1, \cdots, N.$ Then
define the occupation number of $\kappa \in \mathbb{N}$ by
\begin{equation} \label{A1.11} n_\kappa = \sum^N_{j=1} \delta_{\kappa
\kappa_j},
\end{equation}
and the sequence of all $n_\kappa, \kappa \in
\mathbb{N}$, abbreviated $bzf$, by
\begin{equation} \label{A1.12} (n_1, n_2, \cdots) = : \hat n.
\end{equation}
2.) Moreover let us denote the set of all $bzf$, for
$+$ or for $-$, by $BZF$. Then the proposition "$\hat n$ is a sequence
of occupation numbers" is abbreviated by "$\hat n$ is a $bzf$" or by
$"\hat n \in BZF"`$. Sometimes it is useful to write $BZF_L$ for the
set of all $bzf \;\hat l$ with $\sum_\varrho l_\varrho = L$.\\ {\bf
Consequence A.1.6:} 1.) Each sequence of indices $\kappa_1, \cdots,
\kappa_N$ determines uniquely a $bzf$, and vice versa.\\ 2.) The
elements of the ONB \;${\mathcal{B}}^\pm_N$ can be written this way:
\begin{equation} \label{A1.13} \Psi^\pm_{\kappa_1 \ldots \kappa_{N}}
=: \Psi^\pm_N ({\hat n}).
\end{equation}
This notation turns out to be very useful.\\ {\bf
A.1.4:} In the next step the question is to be answered which are the
physically relevant operators, i.e. the relevant observables in
${\mathcal{H}}^N$. Obviously only those are relevant which leave the
spaces ${\mathcal{H}}^N_\pm$ invariant.

Therefore we {\it define}: Let ${\mathcal{D}}_A \subset
{\mathcal{H}}^N$ be the domain of a selfadjoint operator $A$. Then, if
for each $f \in {\mathcal{D}}_A \cap {\mathcal{H}}^N_\pm$ the relation
$A f \in {\mathcal{H}}^N_\pm$ holds, the operator $A$ is called a
physically relevant observable.\\ {\bf Consequence A.1.7:} 1.) $A$ is
physically relevant, exactly if
\begin{equation} \label{A1.14} A = S^+_N A S^+_N + S^-_N A S^-_N +
S^r_N A S^r_N \quad \mbox{with} \quad S^r = 1- S^+_N - S^-_N .
\end{equation}
2.) $A$ is physically relevant, if for each $P \in
{\mathcal{S}} _N$
\begin{equation} \label{A1.15} A= U(P) A U^\star (P)
\end{equation}
holds, i.e. if A is invariant under permutations of
particles.\\ {\bf A.1.5:} Many relevant physical observables are
defined using tensor products of operators in Hilbert spaces.  In
order to avoid unnecessary complications here only the tensor product
of two operators is introduced, because the extension to more than two
factors is straightforward.

Thus let be given two Hilbert spaces ${\mathcal{H}}_1$ and
${\mathcal{H}}_2$ and two densely defined closed operators $A_1$ and
$A_2$. Then $A_1^\star$ and $A^\star_2$ exist having domains
${\mathcal{D}}_{A^\star_1}$ and ${\mathcal{D}}_{A^\star_2}$.

Now form the (noncomplete) tensor product ${\mathcal{D}}_0 : =
{\mathcal{D}}_{A^\star_1} \underline{\otimes }
{\mathcal{D}}_{A^\star_2} \subset {\mathcal{H}}_1 \otimes
{\mathcal{H}}_2$.  It is dense in ${\mathcal{H}}_1 \otimes
{\mathcal{H}}_2 $ and contains only finite linear combinations of the
form
\begin{equation} \label{A1.16} f = \sum a_j \varphi^j_1 \otimes
\varphi^j_2, \quad \varphi^j_\kappa \in {\mathcal{D}}_{A_\kappa},
\kappa = 1,2.
\end{equation}
Then the operator $T_0 (A^\star_1, A^\star_2)$ is
defined by
\begin{equation} \label{A1.17} T_0 (A^\star_1, A^\star_2) f = \sum a_j
(A^\star_1 \varphi^j_1) \otimes (A^\star_2 \varphi^j_2).
\end{equation}
Finally the tensor product of $A_1$ and $A_2$ is {\it
defined} by
\begin{equation} \label{A1.18} A_1 \otimes A_2 = T_0 (A^\star_1,
A^\star_2)^\star.
\end{equation}
Hence, for selfadjoint operators definition
(\ref{A1.18}) reads
$$A_1 \otimes A_2 = T_0  (A_1, A_2) ^\star . $$

A very useful tool is given by \\ {\bf Proposition A.1.8:} For bounded
operators $A_1, A_2$ the above definition of $A_1 \otimes A_2$ is
equivalent to
\begin{equation} \label{A1.19} (A_1 \otimes A_2) g =
\sum^\infty_{\lambda \kappa} a_{\lambda \kappa} (A_1 \phi^1_\lambda)
\otimes (A_2 \phi^2_\kappa)\; ,\\
\end{equation}
where $\{ \phi^\varrho_\lambda : \lambda \in \mathbb{N}
\}$ is an ONB in ${\mathcal{H}}_\varrho, \varrho = 1,2$ and $g =
\sum_{\lambda \kappa}^\infty a_{\lambda \kappa} \phi^1_\lambda \otimes
\phi^2_\kappa \in {\mathcal{H}}_1 \otimes {\mathcal{H}}_2$.\\ {\bf
A.1.6:} The results of the last section now are applied to transfer
observables of $M$-particle systems into observables of $N$-particle
systems, $M<N$. Some basic results in this connection are contained in
the following \\ {\bf Proposition A.1.9:} 1.) Let $A_M$ be selfadjoint
in ${\mathcal{H}}^M$ and let 1 be the identity operator in
${\mathcal{H}}^{N-M}$.Then
\begin{equation} \label{A1.20} (A_M \otimes 1)^\star = A_M \otimes 1
\;.
\end{equation}
2.) If $A_M$ is bounded, then
\begin{equation} \label{A1.21}
\parallel A_M \otimes 1\parallel = \parallel A_M \parallel \;.
\end{equation}
3.) If $(A_M + B_M)^\star = A^\star_M + B^\star_M$,
then
\begin{equation} \label{A1.22} (A_M + B_M) \otimes 1 \supset (A_M
\otimes 1) + (B_M \otimes 1)\; .
\end{equation}
The $=$-sign holds if the domains of both sides are
equal, which is the case if $A_M, B_M$ are bounded and have domain
${\mathcal{H}}^M$.\\ 4.) If $A_M$ or $B_M$ is bounded then
\begin{equation} \label{A1.23} (A_m \otimes 1) (B_M \otimes 1) = (A_M
B_M \otimes 1)\;.
\end{equation}
5.) From (\ref{A1.7}) one concludes that
\begin{equation} \label{A1.24} S^\pm_N (S^\pm_M \otimes 1) = S^\pm_N =
(S^\pm_M \otimes 1) S^\pm_N\;.
\end{equation}
{\bf A.1.7:} The operator which defines the physically
relevant transfer from ${\mathcal{H}}^M$ to ${\mathcal{H}}^N, M < N$
is given by
\begin{equation} \label{A1.25} \Omega_N (A_M) := (M! (N-M)!)^{-1}
\sum_{P \in {\mathcal{S}}_N} U(P) (A_M \otimes 1 \otimes \ldots
\otimes 1) U^\star(P) ,
\end{equation}
where $A_M$ is a densely defined closed linear operator
in ${\mathcal{H}}^M$ and where 1 is the identity operator on
${\mathcal{H}}^1$. It has the following properties. \\ {\bf
Proposition A.1.10:} 1.) The equation
\begin{equation} \label{A1.26} U(Q) \Omega_N (A_M) U^\star (Q) =
\Omega_N (A_M)
\end{equation}
holds for each $Q \in {\mathcal{S}}_N$ so that
$\Omega_N (A_M)$ is indeed physically relevant if it is selfadjoint.\\
2.) If $A_M$ is bounded and selfadjoint with domain ${\mathcal{H}}^M$
the operator $\Omega_N (A_M)$ is bounded and selfadjoint with domain
${\mathcal{H}}^N$.\\ 3.) If $A_M$ is unbounded and selfadjoint then
$\Omega_N (A_M)$ is not necessarily selfadjoint. But it is symmetric
if it is densely defined.\\ {\bf A.1.8:} Though the operators $\Omega_N
(A_M)$ are physically relevant, they are not of interest in a strict
sense, if one wants to consider only systems with one kind of
particles as is the case in this paper. Then only the spaces
${\mathcal{H}}^N_\pm \subset {\mathcal{H}}^N$ are of interest and the
operators defined therein. Thus the following {\it definition} is
natural for the operators of proper physical relevance:
\begin{equation} \label{A1.27} \Omega^\pm_N (A_M) = S^\pm_N \Omega_N
(A_M) S^\pm_N\;.
\end{equation}
As in the preceding sections here $\Omega^\pm_N (A_M)$
is studied only for selfadjoint $A_M$. But the definition itself is
much more general. The operators $\Omega^\pm_N (A_M)$ have some
properties, which are of special interest in this paper.\\ {\bf
Proposition A.1.11:} 1.) For each $P\in {\mathcal{S}}_N $ the following
relations hold:
\begin{equation} \label{A1.28}
\begin{array} {ll} \Omega^\pm_N (A_M) &= {N \choose M} S^\pm_N (A_M
\otimes 1 \cdots \otimes 1)S^\pm_N \\ [3ex] &= {N \choose M} S^\pm_N
U(P) (A_M \otimes 1 \otimes \cdots \otimes 1) U^\star (P) S^\pm_N \; .
\end{array}
\end{equation}
2.) $\Omega^\pm_N (A_M)$ is selfadjoint, because $A_M$
is selfadjoint by supposition and because $S^\pm_N$ are projections,
i.e. are bounded.\\ 3.) If $A_M$ is bounded, then $\Omega^\pm_N (A_M)$
is bounded and
\begin{equation} \label{A1.29}
\parallel \Omega^\pm_N (A_M)\parallel \leq {N \choose M} \parallel A_M
\parallel\; .
\end{equation}
4.) If $A_M + B_M,\; A_M$ and $B_M$ are selfadjoint,
Formula (\ref{A1.22}) implies the relation
\begin{equation} \label{A1.30} \Omega^\pm_M (A_M + B_M) \supset
\Omega^\pm_N (A_M) + \Omega^\pm_N (B_M)\;.
\end{equation}
If $A_M, B_M$ are bounded with domain
${\mathcal{H}}^M$, the $=$ sign holds.\\ 5.) Let $A_M + B_M$ and
$\Omega^\pm_N (A_M) + \Omega^\pm (B_M)$ be selfadjoint. Then
(\ref{A1.30}) is an equation, because a selfadjoint operator cannot
have a selfadjoint extension.\\ 6.)Finally, from (\ref{A1.24}) and (\ref{A1.28})
one concludes that
\begin{equation} \label{A1.31} \Omega^\pm_N (\Omega^\pm_M (A_M)) =
\Omega^\pm_N (A_M)\; .
\end{equation}

\subsection{Dummy Hamiltonians}

{\bf A.2.1:} In order to formulate explicitly the dummy Hamiltonians
for electronic systems it is advisably to work with representations of
Hilbert spaces instead of the abstract versions used elsewhere in this
paper. For our purposes the position-spin representation is most
useful. Therefore the Hilbert spaces we are working with are
\begin{equation} \label{A2.1} {\mathcal{H}}^N_- \subset{\bar
{\mathcal{H}}^N} = \bigotimes^N (L^2(\mathbb{R}^{3}) \otimes
\mathcal{S}^1),\quad {\mathcal{H}}^2_- \subset {\bar {\mathcal{H}}^2}
= \bigotimes^2 (L^2 (\mathbb{R}^{3}) \otimes \mathcal{S}^1) \\ ,
\end{equation}
where $\mathcal{S}^1$ is the complex vector space of
spin functions $u : \{ 1, -1 \} \rightarrow \mathbb{C}$ which is
spanned by the ONB $\{\delta_{1,s}, \delta_{-1, s}\}$.(Cf. also
Subsection 3.1.1.)  This choice fits into the abstract formulations by
the following definition of the tensor product. Let $x \in
{\mathbb{R}}$, $s \in \{1, -1\}$, and $z:= (x,s)$. Then, if $f_1, f_2
\in {\bar{\mathcal{H}}}^1 = L^2 (\mathbb{R}^3) \otimes \mathcal
{S}^1$, one defines $f_1 \otimes f_2$ by
\begin{equation} \label{A2.2} (f_1 \otimes f_2) (z_1, z_2) = f_1 (z_1)
f(z_2)\; .
\end{equation}
For the sake of simplicity let us assume that with the
following two examples the external fields and the interactions are
electrostatic. This means that all influences of magnetism and spin
are disregarded.\\ 
{\bf A2.2:} On the above assumptions the
Hamiltonian for the $N$ electrons in an atom reads
\begin{equation} \label{A2.3} H^-_N \supset \frac{1}{2 m_0}
\sum^N_{j=1} P^2_j - N e_0^2 \sum^N_{j=1} \frac{1}{r_j} + \frac{1}{2}
e^2_0 \sum^N_{j\not= k} \frac{1}{r_{jk}} \;,
\end{equation}
where $r_j = |x_j|$ and $r_{jk} = |x_j -
x_\kappa|$. The domain of $H^-_N$ is ${\mathcal{H}}^N_-$ as defined in
Section 3.1.1. Consequently, the atomic dummy Hamiltonian describes
"dummy helium" and is defined on ${\mathcal{H}}^2_-$. It reads
explicitly
\begin{equation} \label{A2.4} H^-_{20} \supset \frac{\gamma_0}{2m_0}
(P^2_1 + P^2_2) - 2 \gamma_0 e^2_0 (\frac{1}{r_1} + \frac{1}{r_2}) +
e^2_0 \frac{1}{r_{12}}
\end{equation}
with $\gamma_0 = (N-1)^{-1}.$\\ {\bf A2.3:} Now let us
consider $N$ electrons in a finite lattice, the points of which are
given by $y_\alpha, \alpha = 1, \cdots, N$ each carrying the charge
$e_0$. Then the Hamiltonian is defined by
\begin{equation} \label{A2.5} H^-_N \supset \frac{1}{2 m_0}
\sum^N_{j=1} P^2_j - e^2_0 \sum^N_{j=1} \sum^N_{\alpha = 1}
\frac{1}{r'_{j\alpha}} + \frac{1}{2} \sum^N_{j \not= \kappa}
\frac{1}{r_{j \kappa}}\;,
\end{equation}
where $r'_{j \alpha} = |x_j - y_\alpha|$. Thus the
dummy solid has a Hamiltonian given by
\begin{equation} \label{A2.6} H^-_{20} \supset \frac{\gamma_0}{2m_0}
(P^2_1 + P^2_2) - \gamma_0 e^2_0 \sum^N_{\alpha=1}
(\frac{1}{r'_{1\alpha}} + \frac{1}{r'_{2\alpha}}) + e^2_0
\frac{1}{r_{12}}
\end{equation}
and is defined on $\mathcal{H}^2_-.$

\subsection{Hartree-Fock Procedure}

{\bf A.3.1}: In what follows a system of two Fermions is considered. For this purpose it is useful to introduce some {\it notation}.\\
1.) Let $\bar{\mathcal{H}}^1 = L^2 (\mathbb{R}^3) \otimes S^1$, where $S^1$ is the space of spin functions spaned by the $ONB \{\delta_{1S}, \delta_{-1S}\}$. Thus the general two-particle space is $\bar{\mathcal{H}}^2 = \bar{\mathcal{H}}^1 \otimes \bar{\mathcal{H}}^1$ and the space for Fermions is $\bar{\mathcal{H}}^2_-$.\\
2.) It is assumed that the Hamiltonian of the system has the form

\begin{equation} \label{A.3.1}
H_2 = K \otimes 1 + 1 \otimes K + W,
\end{equation} \\
where $W$ is a multiplication operator densely defined in $\bar{\mathcal{H}}^2_-$ by a real function $V (x, s, x', s'), x, x' \in \mathbb{R}^3$ and $s, s' \in \{1, -1\}$. The operator $K$ contains the kinetic energy and the external fields. The dummy Hamiltonian is an example of the operators considered here.\\
3.) The inner product in $\bar{\mathcal{H}}^1$ is defined as usual by 
\begin{equation} \label{A.3.2}
<f, g > = \sum^1_{s = -1} \int \bar{f} (x,s) g(x,s) dx
\end{equation}\\ 
for $f, g \in \bar{\mathcal{H}}^1$.\\
4.) Some  special forms of the inner product appear in the context of Hartree-Fork procedure. Let $\Psi \in \bar{\mathcal{H}}^2$ and $f, g \in \bar{\mathcal{H}}^1$. Then

\begin{equation} \label{A.3.3}
< g, W \Psi >_1 (x,s) = \sum_{s'} \int \bar{g} (x', s') V (x', s', x,s) \Psi (x', s', x, s) dx
\end{equation}
and

\begin{equation} \label{A.3.4}
< g, Wf >_1 (x,s) = \sum_{s'} \int \bar{g} (x', s') V(x', s', x,s) f (x', s') dx'.
\end{equation}\\
Therefore one obtains

\begin{equation} \label{A.3.5}
<g, W (f \otimes h) >_1  =  <g, W f>_1h.
\end{equation}\\
5.) From (\ref{A.3.4}) one concludes that

\begin{equation}  \label{A.3.6}
<g, Wf>_1 = < Wg, f>_1.
\end{equation}\\
If $g=f$, the term $<f, Wf>_1$ is the action of the "charge density" $|f|^2$ on one particle.
Moreover, if the function $V$ is bounded, the term $<g, Wf>_1$ is defined for all $g, f \in \bar{\mathcal{H}}^1$.
For realistic Hamiltonians $\bar H_2$ of the form (\ref{A.3.1}) the function $V$ is symmetric, i.e. $V (x,s,x',s') = V(x', s', x,s)$.\\
{\bf A.3.2:} An essential part of the Hartree-Fock procedure is given by the following\\
{\bf Definition A.3.1}: The linear operator
\begin{equation} \label{A.3.7}
F ( \chi) = K + < \chi, W \chi >_1 - < \chi, W \cdot >_1 \chi
\end{equation}
is defined for all $\chi \in \bar{\mathcal{H}}^1$, for which the domain of $F(\chi)$ is dense in $\bar{\mathcal{H}}^1 \ominus span \{\chi\} = : \bar{\mathcal{H}}^1_\chi$.
It is called the Fock operator belonging to $\chi$.\\
Then for each $\chi$, for which $F(\chi)$ is defined, the following result holds.\\
{\bf Proposition A.3.2:} $F(\chi)$ symmetric in $\bar{\mathcal{H}}^1_\chi$.\\
{\bf Proof:} Let $f,g$ be elements of the domain of $F(\chi)$. Then

\begin{equation} \nonumber
\begin{array} {lll}
<f, F(\chi) g>  
&=&<f, Kg> + < \chi \otimes f, W \chi \otimes g>- <\chi \otimes f, Wg \otimes \chi>\\
&=& <Kf,g> + <W\chi \otimes f, \chi \otimes g> - <W \chi \otimes f, g \otimes \chi>\\
&=& < F(\chi) f, g>^.
\end{array}
\end{equation}\\
{\bf A.3.3:} Now the Hartree-Fock procedure can be described by the following two steps.\\
{\bf 1st step:} Determine two elements $\phi_1, \phi_2$ of $\bar{\mathcal{H}}^1$ and two real numbers $e_{12}$ and $e_{21}$ such that the equations
\begin{equation} \label{A.3.8}
\begin{array} {lll}
F (\phi_2) \phi_1 &=& e_{21} \phi_1 \\

F(\phi_1) \phi_2 &=& e_{12} \phi_2
\end{array}
\end{equation}\\
are satisfied. In addition let $e_{12} \le  e_{21}$.\\
{\bf 2nd step:} Determine normed elements $\phi_\kappa \in \bar{\mathcal{H}}^1_{\phi_1}$, $ \kappa = 3,4, \ldots$ and real numbers $e_{1 \kappa}$ such that the equations
\begin{equation} \label{A.3.9}
F (\phi_1) \phi_\kappa = e_{1 \kappa} \phi_\kappa\\
\end{equation} 
hold.\\
The Hartree-Fock procedure is usually derived via the Ritz variational principle. But this derivation is not of interest in the present context, rather the following consequence.\\
{\bf Proposition A.3.3:} The set ${\mathcal{O}}_1$ of vectors $\phi_\chi \in {\mathcal{H}}^1, \kappa = 1,2,3, \ldots$ obtained from (\ref{A.3.8}) and (\ref{A.3.9}) is an orthonormal system in $\bar{\mathcal{H}}^1$.\\
{\bf Proof:} From the definition of the Fock operator and from the Formulae (\ref{A.3.8}), (\ref{A.3.9}) it follows that $\phi_\kappa \in \bar{\mathcal{H}}^1_{\phi_1}, \kappa \geq 2$. Hence $<\phi_1, \phi_\kappa> = 0$
for all $\kappa \geq 2$. Since the Fock operator $F(\phi_1)$ is a symmetric linear operator in $\bar{\mathcal{H}}^1_{\phi_1}$, all its eigenspaces are orthogonal for different eigenvalues. Therefore the set of eigenvectors $\phi_\kappa, \kappa \geq 2$ can be chosen such that it is an orthonormal system $\bar{\mathcal{H}}^1_{\phi_1}$. Hence, ${\mathcal{O}}_1$ is an orthonormal system in $\bar{\mathcal{H}}^1$.\\
{\bf A.3.4:} The Hartree-Fock procedure is not only a method determining the set ${\mathcal{O}}_1$, rather it is also used for an approximate diagonalization of $\bar H_2$. This runs as follows. Let \\
\begin{equation}\label{A3.10}
\Psi^-_{\kappa \lambda} = \frac{1}{\sqrt{2}}
(\phi_\kappa \otimes \phi_\lambda - \phi_\lambda \otimes \phi_\kappa)
\end{equation}\\
for $\kappa < \lambda $  and let\\
\begin{equation}\label{A3.11}
E_{\kappa \lambda} = <\Psi^-_{\kappa \lambda}, H_2 \Psi^-_{\kappa \lambda}> .
\end{equation}\\
Then the Hartree-Fock approximation of $\bar H_2$ is given by the diagonal operator\\
\begin{equation} \label{A3.12}
\hat{H}^-_2 = \sum_{\kappa \lambda} E_{\kappa \lambda}
\Psi^-_{\kappa \lambda} <\Psi^-_{\kappa \chi}, \cdot >.
\end{equation}\\
Because normally the discret spectrum of $\bar H_2$ is bounded, the operator $
\hat{H}^-_2$ is bounded too. $\hat{H}^-_2$  is the best approximation of $\bar H_2$ using only the Ritz variational principle for vectors of the shape (A3.10).\\
By Proposition 4.1 the Hartree-Fock diagonalization of an N-particle Hamiltonian is reduced to Hartree-Fock diagonalizing its dummy Hamiltonian.

{\bf Acknowledgment}\\ I want to thank my colleagues Arno Schindlmayr, Uwe Gerstmann, Thierry Jecko and J\"org Meyer for valuable discussions and critical
remarks, and Mr. Wolfgang Rothfritz for correcting my English.

\end{document}